\documentclass[sn-aps]{sn-jnl}

\usepackage{graphicx}%
\usepackage{multirow}%
\usepackage{amsmath,amssymb,amsfonts}%
\usepackage{amsthm}%
\usepackage{mathrsfs}%
\usepackage[title]{appendix}%
\usepackage{xcolor}%
\usepackage{textcomp}%
\usepackage{manyfoot}%
\usepackage{booktabs}%
\usepackage{algorithm}%
\usepackage{algorithmicx}%
\usepackage{algpseudocode}%
\usepackage{listings}%

\usepackage{enumitem}
\usepackage{pdflscape}

\begin{document}

\title[Model-Driven Engineering for Machine Learning]{Model-Driven Engineering Method to Support the Formalization of Machine Learning using SysML}


\author*[1,2]{\fnm{Simon} \sur{Raedler}}\email{simon.raedler@tum.de}


\author[1]{\fnm{Juergen} \sur{Mangler}}\email{juergen.mangler@tum.de}
\author[1]{\fnm{Stefanie} \sur{Rinderle-Ma}}\email{stefanie.rinderle-ma@tum.de}

\affil[1]{\orgdiv{TUM School of Computation, Information and Technology; Department of Computer Science}, \orgname{Technical University of Munich}, \orgaddress{\street{Boltzmannstraße 3}, \city{Garching b. München}, \postcode{85748}, \state{Germany}}}

\affil[2]{\orgdiv{Business Informatics Group}, \orgname{Technical University of Vienna}, \orgaddress{\street{Favoritenstraße 9-11/194-3}, \city{Vienna}, \postcode{1040}, \state{Austria}}}

\large

\abstract{\large\textbf{Motivation:} Systems Engineering is a transdisciplinary and integrative approach, that enables the design, integration, and management of complex systems in systems engineering life cycles. In order to use data generated by cyber-physical systems (CPS), systems engineers cooperate with data scientists, to develop customized mechanisms for data extraction, data preparation, and/or data transformation. While interfaces in CPS systems may be generic, data generated for custom applications must be transformed and merged in specific ways so that insights into the data can be interpreted by system engineers or dedicated applications to gain additional insights. To foster efficient cooperation between systems engineers and data scientists, the systems engineers have to provide a fine-grained specification that describes (a) all parts of the CPS, (b) how the CPS might interact, (c) what data is exchanged between them, (d) how the data interrelates, and (e) what are the requirements and goals of the data extraction. A data scientist can then iteratively (including further refinements of the specification) prepare the necessary custom machine-learning models and components. \\
\textbf{Methods:} This work introduces a method supporting the collaborative definition of machine learning tasks by leveraging model-based engineering in the formalization of the systems modeling language SysML. The method supports the identification and integration of various data sources, the required definition of semantic connections between data attributes, and the definition of data processing steps within the machine learning support.\\
\textbf{Results:} By consolidating the knowledge of domain and machine learning experts, a powerful tool to describe machine learning tasks by formalizing knowledge using the systems modeling language SysML is introduced. The method is evaluated based on two use cases, i.e., a smart weather system that allows to predict weather forecasts based on sensor data, and a waste prevention case for 3D printer filament that cancels the printing if the intended result cannot be achieved (image processing). Further, a user study is conducted to gather insights of potential users regarding perceived workload and usability of the elaborated method.\\
\textbf{Conclusion:} Integrating machine learning-specific properties in systems engineering techniques allows non-data scientists to understand formalized knowledge and define specific aspects of a machine learning problem, document knowledge on the data, and to further support data scientists to use the formalized knowledge as input for an implementation using (semi-) automatic code generation. In this respect, this work contributes by consolidating knowledge from various domains and therefore, fosters the integration of machine learning in industry by involving several stakeholders.}




\maketitle
\bmhead{Acknowledgments}
This project has been partially supported and funded by the Austrian Research Promotion Agency (FFG) via the Austrian Competence Center for Digital Production (CDP) under the contract number 881843.

\section{Introduction}
Leveraging data to allow experts making informed decisions during the product lifecycle of a product is recently defined as data-driven engineering \citep{trauer_data-driven_2020}.
The knowledge required for implementing data-driven engineering can be characterized in a two-fold way \citep{hesenius_towards_2019}, i.e., by i) profound machine learning skills with respect to processing and analytics of data and implementation of algorithms, and ii) by domain knowledge regarding the product of interest, relevant product lifecycle data, and related business processes with the entangled IT infrastructures to identify data provenance and information flows.
Regarding i) profound machine learning skills, a recent industrial survey revealed that companies have fewer machine learning experts and too little knowledge to implement solutions themselves. Further, few experts are available on the market \citep{radler_survey_2022}.

To still connect the domain and machine learning knowledge, various methods have been recently proposed in literature \citep{radler_participative_2020,stanula_machine_2018}.
However, these methods lack support for defining machine learning tasks and do not sufficiently represent the perspective of engineers.
Additionally, the methods mainly integrate engineering methods into data science methodologies supporting data scientists rather than allowing engineers to apply the methods to support the elaboration of machine learning support.

Therefore, this work aims to integrate machine learning knowledge into systems engineering to support engineers in the definition of machine learning tasks, to consequently enable data-driven engineering and, ultimately, to support the product development for the definition of prerequisites for the machine learning integration. Particularly, means of Model-Based Engineering (MBE) are adapted to define tasks for data-driven engineering by leveraging data from the product lifecycle of a system.
The method of this work builds upon the systems modeling language SysML \citep{omg_omg_2019}, a general-purpose modeling language allowing to formalize a system from various viewpoints and disciplines. The interdisciplinary formalization of systems knowledge refers to the term Model-Based Systems Engineering (MBSE)~\cite{estefan_survey_2007}. Additionally, the CRISP-DM \citep{wirth_crisp-dm_2000} methodology is used as a basis for the organization of the machine learning task definition. The Cross-Industry Standard Process for Data Mining (CRISP-DM) is a methodology consisting of common approaches used by data mining professionals to work out a data mining project from inception (requirements and business understanding) through processing (data understanding, data preparation and modeling) to evaluation and deployment.
Ultimately, the method proposed in this work aims to formalize machine learning tasks during product development and to use the formalized knowledge to derive parts of the machine learning and to guide the implementation, respectively. The method is evaluated using a case study representing a weather station with multiple subsystems to predict weather forecasts and a second study to prevent wasting of 3D printer filament by canceling the printing if the intended result cannot be achieved.

The contribution of this work is manifold:

\begin{itemize}
  \item The proposal of a SysML metamodel extension to include stereotypes that are used to describe machine learning functions for domain-specific data objects
  \item A method that fits to the latest research areas of the modeling community and is called MDE4AI~\cite{burgueno_mde_2021, burgueno_mde_2021-1}
  \item A means of structuring the models based on the CRISP-DM methodology.
  \item Two case studies using the proposed concepts for modeling machine learning support based on simple input data, followed by a discussion of the strengths and weaknesses of the method.
  \item A user study showing the workload and usability of the method as rated by experts and computer scientists.
\end{itemize}

This work lays a foundation for allowing non-programmers to define machine learning tasks by formalizing knowledge from the problem domain into a high-level model and to communicate formalized knowledge.
Additionally, semantic connection of data from various Product-Lifecycle Management (PLM) \citep{stark_plm_2016} sources allows to describe the origination and composition of data relations.
With the availability of such models, the goal is to support the automatic decomposition of SysML models and the (semi-)automatic generation of executable machine learning modules.

This work constitutes an extension of our previous work presented in 
 \cite{radler_integration_2022} and expands \cite{radler_integration_2022} in several ways by
\begin{itemize}
    \item providing more extensive background information to foster understanding.
	\item extending the presented method with a generic and fine-grained sample of the modeling method.
	\item applying the method in two case studies from industry.
    \item conducting a user study on the perceived workload and usability of mechanical engineers and computer scientists
	\item discussing advantages and disadvantages of the method in a more thorough way.
\end{itemize}

The remainder of this paper is structured as follows: Section \ref{sec:backgroundCRISP} presents the background regarding MBSE, data science methodologies and related work of data-driven engineering. In Section \ref{sec:method}, the elaborated method is introduced in detail and evaluated based on two case studies in Section \ref{sec:casestudy}. Further, a user study is presented in Section \ref{sec:userstudy} that evaluates the perceived workload and the usability of the method with mechanical engineers and computer scientists.
Based on the findings of the evaluation and the user study, an extensive discussion on advantages and disadvantages is presented in Section \ref{sec:discussion}. Finally, the study is summarized in conclusion with future remarks in Section \ref{sec:conclusion}.

\section{Background}
First, the concepts of model-based systems engineering (MBSE) and the systems modeling language SysML are explained. Second, machine learning and the CRISP-DM \citep{wirth_crisp-dm_2000} methodology are introduced, acting as a basis for the method presented in Section \ref{sec:method}. Next, related methods are depicted with special focus on data-driven engineering. Finally, Section \ref{sec:summary} presents a summary of the background.

\subsection{Model-Based Systems Engineering and SysML}\label{sec:backgroundSysML}
\begin{figure}
    \centering
    \includegraphics[width=0.8\linewidth]{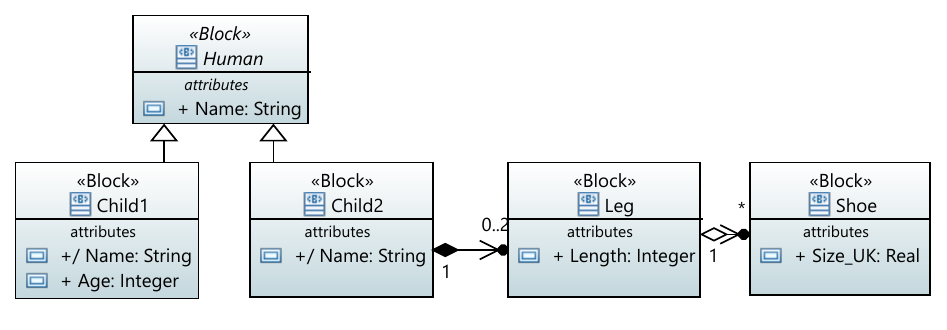}
    \caption{Block Definition Diagram sample with a human.}
    \label{fig:human}
\end{figure}

Systems engineering, particularly MBSE, aims to integrate various engineering disciplines in product development to establish a single-source of truth by formalizing system requirements, behavior, structure and parametric relations of a system. Conventional systems engineering focuses on storing artifacts in several (text) documents maintained in case of changes. In a model-based method, the relevant information to describe an abstract system are stored in a model \citep{madni_model-based_2018}. The literature concerning graphical MBSE methods promises to increase design performance while supporting the communication of relevant stakeholders of a system \citep{huldt_state--practice_2019,henderson_value_2021}.
MBSE is a term explicitly considering aspects of a system. Nevertheless, other terms can be considered interchangeable depending on the level of automation and the focus of the application\footnote{See \url{https://modelling-languages.com/clarifying-concepts-mbe-vs-mde-vs-mdd-vs-mda/} for a discussion.}. Independent of the level of automation and the focus of the modeling language, a metamodel defines the modeling concept, relations and all possible instances of a specific set of models. Models are instances of metamodels describing a specific system. The model characteristics must match all aspects of the associated metamodel. However, extensions such as additional attributes can be added directly on a model without changing the metamodel. If a metamodel does not represent an aspect, an extension for a specific group of use cases can be defined using so-called stereotypes \citep{brambilla_model-driven_2017}. A stereotype is a means of modeling to extend metaclasses by defining additional semantics for a specific class concept. A metaclass is a class describing a set of classes, e.g. the metaclass \textit{block} is a general purpose structuring mechanism that describes a system, subsystem, logical or physical component without the software-specific details implicitly given in UML structured classes \citep{omg_omg_2019}.
The use of stereotypes in modeling methods have been proven to support the understanding and standardization of a model \citep{kuzniarz_empirical_2004}.
In MBSE, the Systems Modeling Language SysML is the most prominent modeling language \citep{albers_challenges_2013}. SysML is based on the UML standard with a special focus on the formalization of systems instead of modeling classes and objects for software engineering. The language supports the formalization of structural, behavioral and functional specifications \citep{holt_sysml_2013}.
Structural diagrams describe the composition of systems and subsystems with their attributes and relations \citep{holt_sysml_2013,brambilla_model-driven_2017}. Figure \ref{fig:human} depicts core elements of a block definition diagram modeled in the Eclipse-based open-source software Papyrus\footnote{\url{https://www.eclipse.org/papyrus/index.php}}. On top of \autoref{fig:human}, a Block with the name \textit{Human} is defined, consisting of one attribute of type \textit{String} with the attribute name \textit{Name} and the visibility \textit{public} indicated by the plus (+). A block can also have operations, ports etc. which are not relevant for this work and, therefore not introduced here. Underneath the \textit{Human}-Block, two inheriting elements are defined by the white arrows between the blocks. The attribute \textit{Name} is inherited from the parent block marked by the tailing dash. One child has an additional property \textit{Age}, which only affects the block (as long as no deeper inheritance is available). The second block consists of a subsystem, indicated by the black diamond being a part association (a.k.a. composition). A part association determines that a block describes a whole element and a part of the whole element is additionally described in another element\footnote{See \url{https://sysmlforum.com/sysml-faq/what-are-diff-among-part-shared-referenced-associations.html} for a discussion}. The 1 and the 0..2 indicate the multiplicity, allowing to define the cardinality, e.g. number of elements. In this sample, it means one element \textit{Child2} can have zero, one or two legs. The white diamond between \textit{Leg} and \textit{Shoe} indicates a shared association, which is a weaker form of the part association. It refers to a relationship where the part element is still valid if the whole element is deleted, e.g. if the element \textit{Leg} is not valid anymore, the \textit{Shoe} is still valid. The multiplicity * indicates that one can have any number of shoes.
Since various software represent slightly different parts, the description of the block definition diagram can vary.

\begin{figure}
    \centering
    \includegraphics[width=0.6\linewidth]{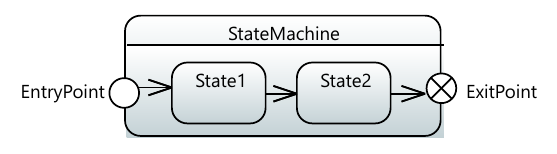}
    \caption{State diagram sample.}
    \label{fig:human-state}
\end{figure}

In SysML, the execution of single activities can be modeled using activity diagrams. A state diagram has an entry-point and an exit-point. The arrow between the states indicates a transition and describes that one state has been completed and another is active. Behind a state, the execution of one or multiple activities can be triggered, whereas an activity is a sequential execution of single actions \citep{omg_omg_2019}, see \autoref{fig:human-state}.

\subsection{Data Science and Methodologies}\label{sec:backgroundCRISP}
Data Science and Business Intelligence refer to the extraction of information and knowledge from data through analysis to assist people with various types of insights, such as analysis or prediction, among many others \citep{sanchez-pinto_big_2018, grossmann_fundamentals_2015}.
The digging of such information to derive knowledge is called data mining (DM)\citep{provost_data_2013}.
Machine learning (ML) is one subfield of DM, which automatically allows computer programs to improve through experience \citep{carbonell_overview_1983}.
Machine learning algorithms aim to solve a (specific) problem to eliminate the need for being explicitly programmed \citep{koza_automated_1998}.

To support the implementation of machine learning applications, methodologies have been proposed in a general manner \citep{wirth_crisp-dm_2000,shafique_comparative_2014,fayyad_kdd_1996}. Additionally, extensions of such methods with particular support for data science in the engineering domain are introduced \citep{radler_participative_2020,bitrus_applying_2021}.
In literature, the methods of the CIRSP-DM \citep{wirth_crisp-dm_2000} and KDD~\citep{fayyad_kdd_1996} are assessed in a comparative study \citep{azevedo_kdd_2008}. According to \cite{azevedo_kdd_2008}, CRISP-DM is a kind of implementation of the KDD process. In the following, CRISP-DM is described and used as basis for the structure of the proposed method described in Section \ref{sec:method}.

In CRISP-DM, six core steps are defined supporting the implementation of a DM application:
\begin{enumerate}
    \item \textbf{Business Understanding: }Project objectives, requirements and an understanding from a business level is achieved. Based thereon, a DM problem is defined and a rough roadmap is elaborated.
    \item \textbf{Data Understanding: }Data is collected to understand the situation from a data point of view. 
    \item \textbf{Data Preparation }The construction of the final dataset for the learning algorithm based on raw data and data transformations. 
    \item \textbf{Modeling: }Various or sometimes one algorithm is selected and applied to elaborated dataset from the previous step. In this step, so-called hyperparameter tuning is applied to vary on parameter values and achieve a most valuable result. 
    \item \textbf{Evaluation: }The result of the algorithm is evaluated against metrics and the objectives from the first step. 
    \item \textbf{Deployment: }The achievements are presented in a way that a customer or an implementation team can use it for further integration.
\end{enumerate}

\subsection{Related Work}
In literature, various methods supporting the formalization of data-driven engineering or machine learning using modeling languages, are given.
The method of \cite{hartmann_next_2017} is based on the Kevoree Modeling Framework KMF~\citep{fouquet_eclipse_2012}, which is similar to the Eclipse Modeling Framework (EMF) that is the basis for the open source modeling framework Papyrus\footnote{\url{https://www.eclipse.org/papyrus/}}. \cite{hartmann_next_2017} proposes to model the domain knowledge and small learning units in a single domain modeling method since both are highly entangled. The method is based on a textual modeling syntax and describes what should be learned, how and from which attributes and relations. Additionally, templates are given to render code based on the model. However, the open-source framework seems to be out of maintenance since the repository is not updated since 2017\footnote{\url{https://github.com/dukeboard/kevoree-modeling-framework}}.

An active maintained framework family with means to model machine learning is shown in \cite{kusmenko_engineering_2019}. The method is based on the MontiAnna framework \citep{kusmenko_modeling_2019} and focuses on modeling artificial neural networks. The MoniAnna framework is part of the MontiCore Workbench Family\cite{rumpe_monticore_2017}.
Similar to \cite{hartmann_next_2017}, textual modeling is used to formalize the learning units and related input and output. The formalization is used as input for template-based code generation. However, the method does not reflect domain-specific (business) knowledge from an engineering perspective.

In \cite{bhattacharjee_stratum_2019}, focus is put on the integration of executable machine learning units modeled on a cloud platform, enabling the fast deployment of distributed systems. However, the method is stiff regarding extendability and advanced data preparation as of the current development state.Additionally, the integration of domain knowledge is hardly given and the focus on the formalisation of data-driven algorithms is not present.

The integration of ML in CPS modeling is supported by the textual modeling framework ThingML+\citep{moin_model-driven_2022}. The method extends the ThingML \citep{harrand_thingml_2016} modeling method, intended to support the development of IoT devices. As with the other methods, focus is put on machine learning modeling without considering domain knowledge. The method allows deriving executable code based on model transformation using xtext.

\subsection{Summary}\label{sec:summary}
MBSE has been proven beneficial in increasing the design performance of systems \citep{huldt_state--practice_2019,henderson_value_2021}. According to \cite{beihoff_world_2014}, the number of components and functions are increasing in future, leading to more complex systems, requiring advanced support in the development and analysis using means of data science.
Development support for data science is given in methodologies such as CRISP-DM. However, guidance specific for the engineering domain is limited \citep{bitrus_applying_2021} and the integration in a model-based method is unavailable as of the author's knowledge.
In literature, various methods introduce specific metamodels and languages to describe a data science task and eventually enable to derive executable code. However, the methods are not based on a MBSE compatible modeling language such as SysML rather than introducing single domain-specific modeling environments.
Therefore, little support for interdisciplinary communication is given and the methods are more applicable for computer scientists than to domain outsiders such as mechanical engineers with little knowledge in programming. Moreover, the domain-specific modeling methods are not aligned with the CRISP-DM methodology, leading to little support from a methodological perspective. Last but not least, the proposed methods use model transformation to reduce the implementation effort, but are seldomly built in a generic way, allowing to extend the modeling or the derivation of code without extensive changes in the generation. Therefore, maintenance and applicability in practice is rather limited.




\section{Method} \label{sec:method}
This section describes a method to formalize machine learning tasks based on SysML and the application of an extended metamodel.
In the following, first, the extension of the SysML metamodel using stereotypes is described.
Special attention is given to the package structure for organizing the stereotypes, extensibility for different purposes, and generalization so that stereotypes can be used for multiple use cases. 
Second, a package structure aligned with the CRISP-DM methodology is presented, enabling to guide the application of the newly defined stereotypes.
Next, a syntax and semantic is introduced, allowing to interpret the formalized machine learning model enriched with the introduced stereotypes.
Finally, means of SysML state diagram is used to define the tasks' execution order.

\subsection{Metamodel Extension using Stereotypes} \label{sec:metamodel}

In the following subsections, six packages are introduced, which allow to group stereotypes that semantically describe required functionalities.
Subsequently, an exemplary stereotype hierarchy for defining higher-order functions for domain-specific data transformation purposes is described in detail.

\subsubsection{Stereotype Package Structure}\label{sec:StereotypePackages}

\begin{figure}
    \centering
    \includegraphics[width=0.75\linewidth]{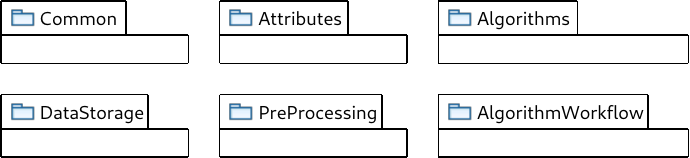}
    \caption{The organization of the metamodel.}
    \label{fig:packages}
\end{figure}

SysML packages are used to group and organize a model and to reduce the complexity of system parts.
Similarly, it can be applied for the organization of stereotypes, as depicted in Figure \ref{fig:packages}.

The organization of the stereotypes is as follows: in \textit{Common}, general stereotypes are defined that are used in other packages as basis, e.g. a stereotype \textit{ML} is defined in \textit{Common}, each defined stereotype related to machine learning inherits from this stereotype to indicate that it is a machine learning stereotype. Additionally, stereotypes can be defined allowing to categorize other stereotypes, e.g. an abstract \textit{Pre-Processing} stereotype allows to identify that all inheriting stereotypes are introduced for the data preparation step of the CRISP-DM methodology.
In \textit{Attributes}, stereotypes for a more detailed definition of attributes are defined. These attribute stereotypes cannot be applied to blocks, only to attributes of a block. Consequently, the stereotypes extend primitive data types such as \textit{Integer} or \textit{Float}. The purpose of the extension are additional characteristics to describe the data, e.g. valid ranges of a value or the format of a datetime property or a regular expression to collect or describe a part of a text value.
The package \textit{DataStorage} defines available data interfaces from a general perspective required for the loading and processing of data from various data sources, e.g. SQL servers, Application Programmable Interface (API) or other file formats (e.g. CSV). 
The purpose of the stereotypes are to support the \textit{data understanding} of the CRISP-DM methodology. Additionally, it allows to bridge the gap between business and data understanding due to the explicit formats. Further details in Section \ref{sec:semantics}.
In the \textit{Algorithm} package, various machine learning algorithms are defined and grouped with respect to algorithm types, e.g. regression or clustering algorithms. Particularly, the focus is put on key characteristics of an algorithm implementation, such as mandatory hyper-parameter or the stereotype description. Optional algorithm parameters are not described in the stereotype, but can be added during the modeling, as later illustrated in Figure \ref{fig:implPreProcessing}.
The \textit{PreProcessing} package (a.k.a. as data preparation) is the most complex and extensive package due to the number of functionalities required. Additionally, a survey revealed that computer scientists spend the most effort in preparing and cleaning data \citep{anaconda_state_2022}. Within this package, functions are defined allowing to transform data so that a cleaned and applicable dataset for the machine learning algorithm is defined.
Finally, the \textit{AlgorithmWorkflow} package, consisting of stereotypes for states of the state diagram, allowing to define the implementation order of the machine learning tasks. Typically in SysML, states are connected to activities, which are a sequence of execution steps. However, in practice, we found out that it is very time consuming to prepare activities first. Additionally, a function abstracted as a single block can be considered as a set of activities. Consequently, state diagrams are used instead of activity diagrams to reduce the implementation effort and complexity.

\subsubsection{Stereotypes Hierarchy}
\begin{figure}
    \centering
    \includegraphics[width=\linewidth]{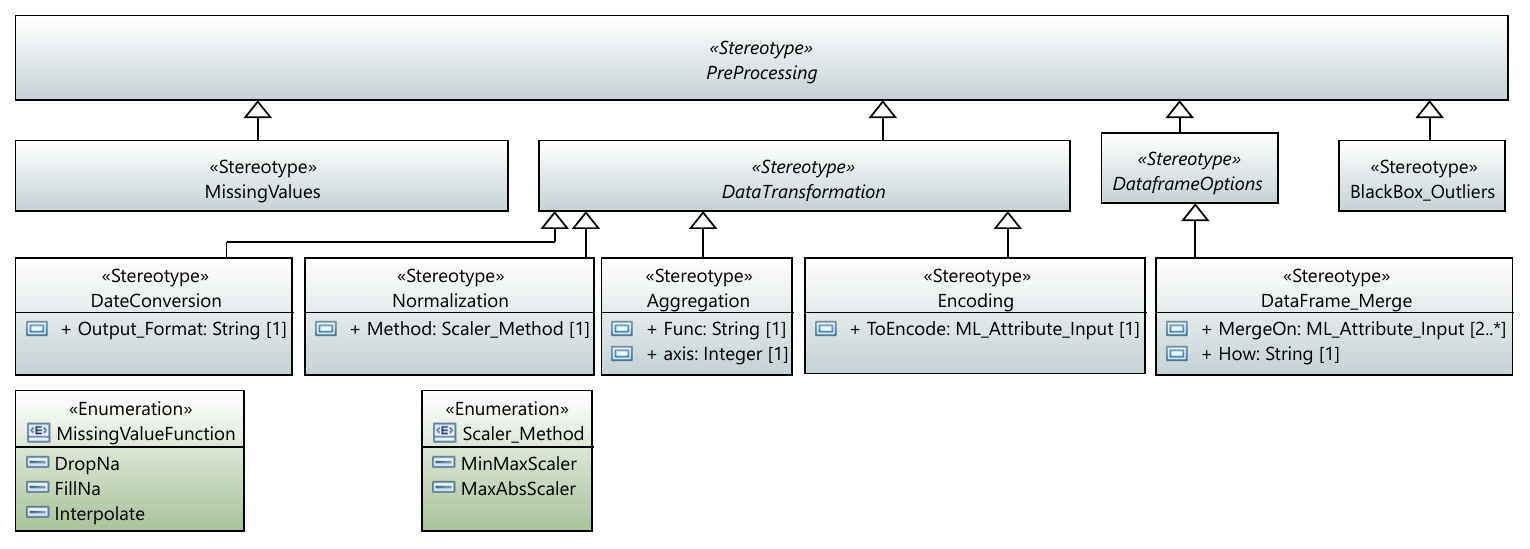}
    \caption{The metamodels for data pre-processing/preparation.}
    \label{fig:ml-preprocessing}
\end{figure}

As mentioned in Section \ref{sec:StereotypePackages}, each package represents a specific hierarchy of stereotypes, allowing to describe various aspects of machine learning subtasks. 
An example definition of stereotypes related to data pre-processing is depicted in Figure \ref{fig:ml-preprocessing}. As described in Section \ref{sec:backgroundSysML}, stereotypes can be hierarchically composed to describe specific attributes only once for a set of stereotypes.
On top, the \textit{ML} stereotype defined in the \textit{Common} package is depicted, indicating that all inheriting stereotypes are related to machine learning. Formalizing a machine learning task is intended to be iteratively, which is why some stereotypes are abstract, illustrated by italic letters. 
If a stereotype is abstract, it means that the stereotype requires further detailization or that a child stereotype with additional information is required, e.g., \textit{DataTransformation} cannot be used without further details as it can be arbitrary transformation of data. 
The purpose of abstraction is to support the early definition of tasks in the product development without details already being known, e.g., the final file-format used to store the data. 
From top to bottom in Figure \ref{fig:ml-preprocessing}, the level of detail increases and the task is more fine-grained chosen. Consequently, leaves are the most fine-grain representation. The inheritance additionally allows to group functions of a specific kind, e.g., functions regarding outlier detection etc.
Due to the grouping of functions, the composition of stereotypes strongly depends on the preferences of the implementing expert and the purpose of the composition in terms of inheritance of attributes.
Note that attributes defined in a parent stereotype are also available in a child or grandchild stereotype, respectively. Therefore, each level should only represent mandatory attributes. This especially applies for algorithms with a lot of hyper-parameters, e.g. logistic regression with more than 20 parameter and attributes\footnote{\url{https://scikit-learn.org/stable/modules/generated/sklearn.linear_model.LogisticRegression.html}}. In case a parameter is not defined in the stereotype, it sill can be add during the modeling and application of the stereotypes. A sample can be found in Section \ref{sec:casestudy}. 
Additionally, it is possible to add a set of values using \textit{Enumerations} for a single attribute, e.g. \textit{MissingValueFunction} highlighted in green. In this respect, modeling is more precise and guided by a fixed set of valid options.
Similarly, specific stereotypes can be used as an attribute, which means that only blocks or attributes that apply the specific stereotype can be assigned, e.g. \textit{Method\_Attribute\_Input} indicating that only properties with a stereotype defined in the package \textit{Attributes} can be applied because each attribute stereotype inherit from that stereotype.
Finally, the application of the keyword \textit{BlackBox} can be used if a function shall be hidden due to security reasons or the implementation is unknown, e.g. \textit{BlackBox\_Outliers} on the right side of Figure \ref{fig:ml-preprocessing}.

\subsection{Package structure guiding the implementation.}

\begin{figure}
    \centering
    \includegraphics[width=0.75\linewidth]{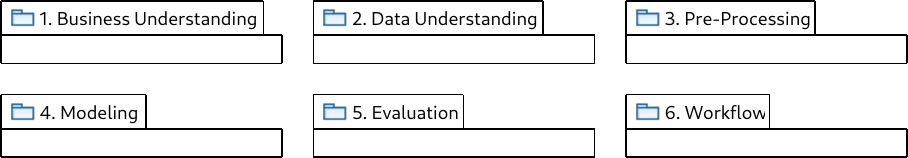}
    \caption{The implementation structure aligned with CRISP-DM.}
    \label{fig:implStructure}
\end{figure}

CRISP-DM as described in Section \ref{sec:backgroundCRISP} consists of six steps, each describing a specific aspect required for the development of a machine learning project. Figure \ref{fig:implStructure} illustrates the package structure aligned with the CRISP-DM methodology. 
\textit{Business Understanding} consists of block definition diagrams describing the system under study with the composition from a system configuration point of view. In this respect, the VAMOS method (Variant Modeling with SysML, \cite{weilkiens_variant_2014}) is integrated to describe a specific system configuration. The integration of the VAMOS method focuses on the data interfaces and attributes of a particular configuration of a system, as different configurations of a system might lead to other data output. In this method, the VAMOS method is used to focus on data interfaces. Therefore, other systems engineering knowledge is presented in other diagrams, which is out of the scope of this work. Still, the knowledge modeled in other diagrams is connected to the instance of a block used in the VAMOS method and therefore, multiple disciplines are enabled to work on the same model.
The second step, \textit{Data Understanding}, details the \textit{Business Understanding} with the definition of delivered data on an attribute and data format level. Particularly, the data type and the name of the delivered data attribute are described using block definition diagrams. Additionally, attribute stereotypes are used to describe the data in detail as described in Section \ref{sec:StereotypePackages}. With the application of stereotypes on a block level, the type of data interface is defined, e.g. CSV files or SQL servers. As a result of the formalization of the interfaces in this package: The information exchange between the systems engineering and the data engineering can be considered as completed.
Based on the \textit{Data Understanding}, the \textit{Pre-Processing} is applied to transform and prepare the data in a final dataset that can be used in the \textit{Modeling}. In the \textit{Pre-Processing}, the most effort is required due to the possible number of required data transformations to create a dataset usable for machine learning. The result of the \textit{Pre-Processing} is a final dataset, considered to be ready for the machine learning algorithm.
Within the \textit{Modeling}, algorithms are applied to the final dataset. Additionally, train-test-splitting and other required functions on the machine learning algorithm are applied.
In the \textit{Evaluation} package, various metrics are used to asses and prove the validity of the algorithm result of the \textit{Modeling} package.
Finally, the \textit{Workflow} package, which describes the execution order of the formalization in the previous packages using state diagrams. For each state, a custom stereotype is applied allowing to connect a block that is connected to a stereotype inherited from \textit{ML}. The method to assign blocks to states allows to overcome the necessity to define activities, making the method less heavy for the application and reduces time for the formalization of the machine learning.
Typically in CRISP-DM, the very last step is the \textit{deployment}. However, the deployment is considered out of scope in this work and therefore the method ends with the workflow.

\subsection{Syntax and Semantics}\label{sec:semantics}
For the purpose of implementing ML functionalities, the utilization of
functional programming paradigm is intuitive \citep{nygaard_basic_1986}. It
utilizes higher order functions, invoked on (data-)objects which are returning
objects. This allows for step-by-step decomposition, filtering and
transformation of data, without side-effects (changes to variables), in
comparison to the imperative programming paradigm.

This sequence of function invocation aligns well with how UML and other
modeling languages implement abstraction-levels to reflect a relevant selection
of properties to focus on the aspects of interest
\citep{brambilla_model-driven_2017}. Functions are blackboxes with processing
capability that are associated with (data-)artifacts upon which they can be
called, and are associated with data-artifacts they produce as output. The
abstraction is realized by describing functions or a set of functions with a
single stereotype and instances with blocks. \newline

\begin{figure}
    \centering
    \includegraphics[width=0.75\linewidth]{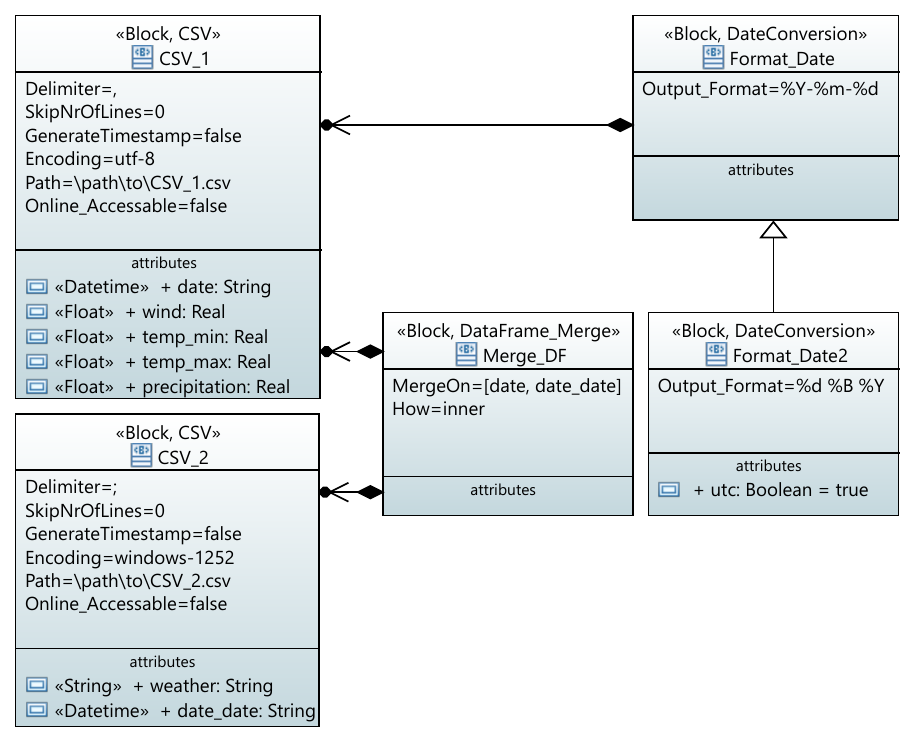}
    \caption{Machine learning data pre-processing based on a sample in Section \ref{sec:casestudy}.}
    \label{fig:implPreProcessing}
\end{figure}

A class in UML is defined among others by attributes, stereotypes, operations (methods), constraints and relationships to other classes. In SysML, a block describes a system or subsystem with a similar definition as a class in UML. A machine learning task and the respective subtasks can be seen as a system with subsystems. Therefore, each subtask is modeled using blocks, aligned with the syntax described in section \ref{sec:backgroundSysML}. Particularly, only input values represented as attributes of a block and the relation to other blocks are modeled. The operations (methods) are defined as stereotypes with abstracted implementations. Attributes defined on the stereotype are mandatory input values for the definition of a machine learning subtask. The attributes defined on a block itself are optional for documentation or to extend the stereotype with fine-grained details, e.g. \textit{utc} attribute in the \textit{Format\_Date2} block in Figure \ref{fig:implPreProcessing}. The output of a subtask (block) is implicitly defined in the implementation of the code snippet related to a stereotype and not explicitly depicted in the model. The output of a block can be used as input for other blocks, e.g. \textit{CSV\_1} block as input for the \textit{Format\_Date} block. \newline
Figure \ref{fig:implPreProcessing} depicts a few samples of the aforementioned syntax and semantics. On top right, a date conversion subtask is modeled as \textit{Format\_Date}. The date conversion stereotype has a mandatory attribute to define the format of the output of the conversion. The input for the date conversion is the block \textit{CSV\_1}, connected using a part association. In this sample, the \textit{date} attribute is the only input value matching due to the stereotype \textit{Datetime}. However, if the input is ambiguous because the datetime is stored for instance as integer or multiple attributes of the connected block are in the correct input format, it is necessary to add additional attributes to the date conversion to select the particular input, e.g. with a new attribute which value is the particular input attribute from the connected block.
The block \textit{Format\_Date2} inherits from \textit{Format\_Date}. Therefore, the input and the attributes are the same except of manual overwritten values, e.g. changes on the output datetime format or the added additional attribute \textit{utc}. \newline
Another sample in Figure \ref{fig:implPreProcessing} shows the integration of multiple inputs. The \textit{Merge\_DF} block consists of two input blocks and the attributes on which the merging function shall be applied are defined using an attribute that consists of two values (\textit{MergeOn}). The \textit{MergeOn} attribute is mandatory and therefore defined on the stereotype.\newline
Although the implicit execution order of the subtasks is defined by the associations and the necessity to compute first inputs, the execution order might be ambiguous, e.g. execute first the \textit{Format\_Date} or the \textit{Merge\_DF}. As described in section \ref{sec:backgroundSysML}, structural diagram elements, such as blocks, requires the integration in behavioral diagrams to allow the definition of an execution order \citep{brambilla_model-driven_2017}.\newline
To enable the connection of a block with a state in a state diagram, custom stereotypes are applied. The stereotypes for the states consist of a single mandatory attribute. The mandatory attribute references a block with a stereotype that inheritate from the root parent stereotype \textit{ML}.

\section{Case Studies}\label{sec:casestudy}
This section presents two case studies, i.e., a weather system that predicts weather forecasts based on sensor data, and an image similarity check that makes it possible to assess whether the actual print of a 3D model with a 3D printer corresponds to the desired output.
As a result, the printing process can be stopped prematurely, saving filament and time.

\subsection{UC1 - Weather Forecast based on Sensor Data}
\begin{figure}
    \centering
    \includegraphics[width=0.85\textwidth]{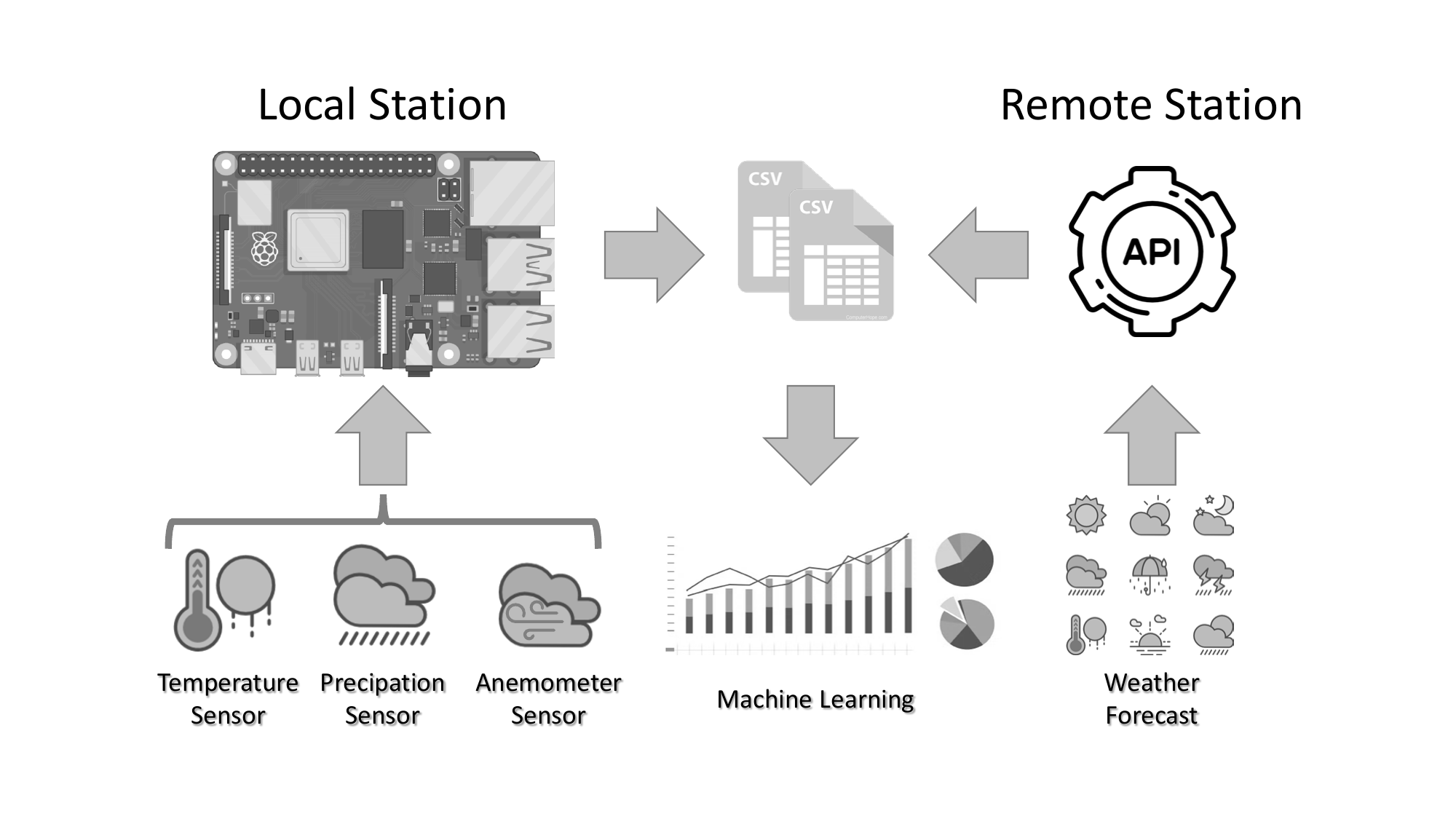}
    \caption{Illustration of the weather system use case.}
    \label{fig:WeatherUseCase}
\end{figure}

Figure \ref{fig:WeatherUseCase} illustrates the composition of the weather system that is split in two parts. On the left side, a local station is equipped with various sensors, delivering a CSV file with measuring and on the right side, a weather forecast additionally delivers a CSV file with weather forecasts over the internet.

\begin{figure}
    \centering
    \includegraphics[width=0.7\textwidth]{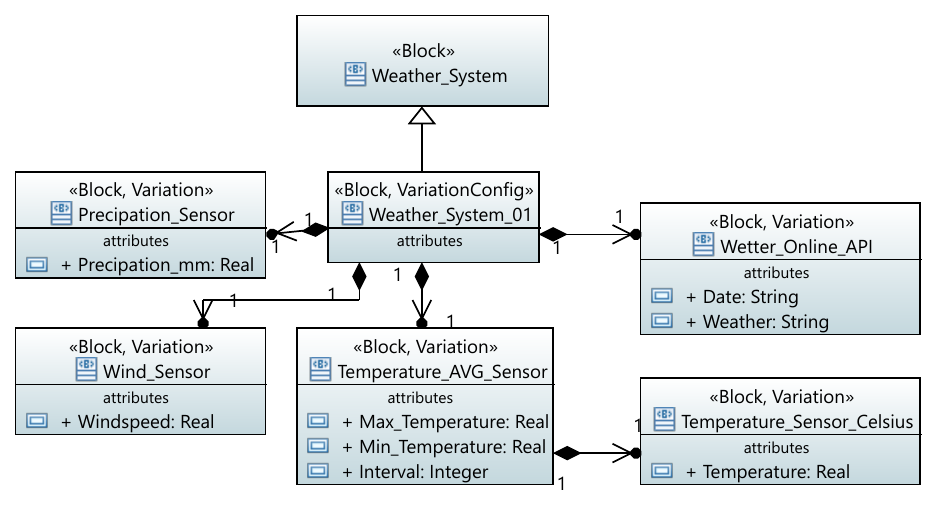}
    \caption{Business Understanding of the weather system.}
    \label{fig:weatherbusiness}
\end{figure}

From a systems engineering perspective, the weather system is a cyber-physical system and can be configured with various sensors.
Figure \ref{fig:weatherbusiness} depicts the SysML model of the weather system with a specific configuration aligned with Figure \ref{fig:WeatherUseCase}.
Particularly, Figure \ref{fig:weatherbusiness} depicts an method aligned with \cite{weilkiens_variant_2014} that allows to formalize variations. Additionally, the modeling of the system from an business perspective is the first step of the method. Focus is put on the values of interest, which are the output values of the subsystems, to keep the business understanding as concise as possible. In the middle of the figure, the core weather system configuration is depicted. The surrounding subsystems are sensors or subsystems, e.g., an API (right side). The attributes of the sensors are output values of each subsystems to align with the CRISP-DM business understanding that aims to get a general idea of the system and from where data originates.

To transform the business understanding in valuable data understanding, connections between the system in the business understanding and output data formats are established.
Particularly, a \textit{realization} connection between the CPS and blocks describing the data format using stereotypes inheriting from \textit{ML} are modeled.
In the blocks, each attribute has a type representing the actual data type in the data source and a stereotype with a \textit{ML} attribute describing the representation in the machine learning method, e.g., \textit{CSV\_2} attribute \textit{date\_date} is of type \textit{String} and is mapped to the stereotype \textit{Datetime} that considers aspects such as the datetime format.
Additionally, stereotype attributes are defined such as the \textit{Encoding} or the \textit{Delimiter} to describe the composition of the \textit{CSV} file.

\begin{figure}
    \centering
    \includegraphics[width=0.5\textwidth]{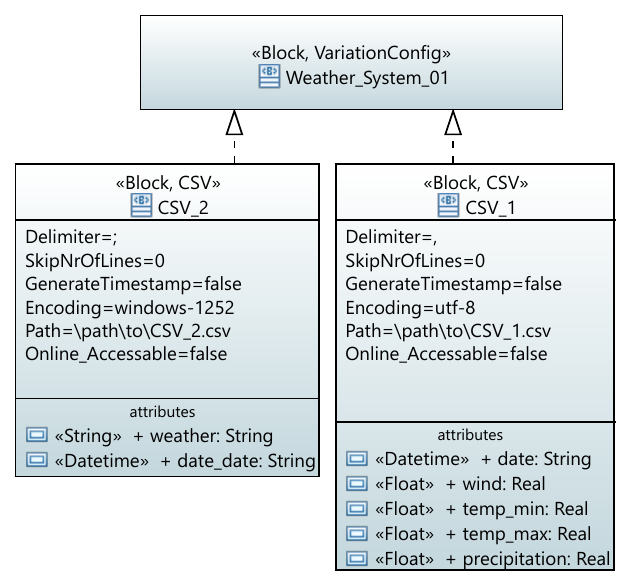}
    \caption{Data Understanding of the weather system.}
    \label{fig:weatherdata}
\end{figure}

Figure \ref{fig:implPreProcessing} depicts a set of subtasks applied to the data sources defined in Figure \ref{fig:weatherdata}.
For and explanation of Figure \ref{fig:implPreProcessing}, please refer to Section \ref{sec:method}.

\begin{figure}
    \centering
    \includegraphics[width=0.8\textwidth]{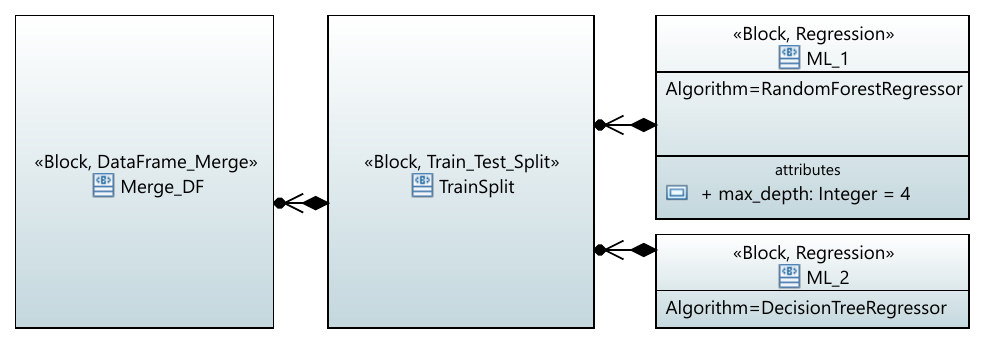}
    \caption{Modeling of machine learning algorithms.}
    \label{fig:weathermodeling}
\end{figure}

Figure \ref{fig:weathermodeling} illustrates the application of a train-test-split and the integration of the split data into two different regression algorithms, which are specified in a mandatory attribute. As of the definition of the stereotypes, no further parameters are mandatory. For the \textit{RandomForestRegressor}, the optional hyper-parameter \textit{max\_depth} is defined.

\begin{figure}
    \centering
    \includegraphics[width=0.7\textwidth]{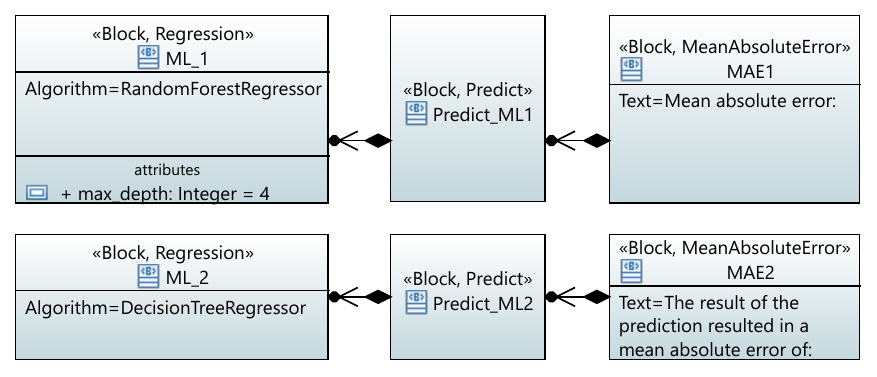}
    \caption{Evaluation of the weather forecast prediction.}
    \label{fig:weatherevaluation}
\end{figure}

Figure \ref{fig:weatherevaluation} depicts the prediction and the application of metrics such as mean absolute error (MAE). The mandatory parameter text is a placeholder allowing to add text that shall be implemented with the evaluation result.

\begin{figure}
    \centering
    \includegraphics[width=0.9\textwidth]{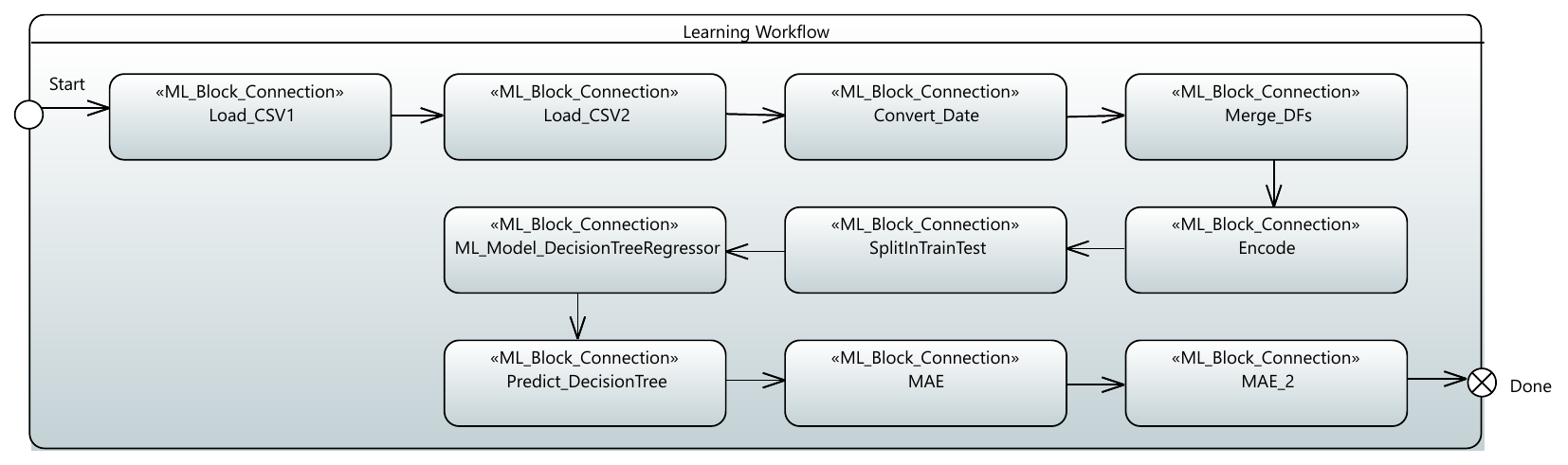}
    \caption{Sample integration of the workflow.}
    \label{fig:weatherworkflow}
\end{figure}

The method's final step is integrating the blocks into an execution workflow. Figure \ref{fig:weatherworkflow} illustrates the execution order of the algorithm steps. As can be seen, the \textit{Format\_Date2} block modeled in Figure \ref{fig:implPreProcessing} is not depicted in the workflow, meaning that it is not taken into concern during the implementation and is left out as an artifact from the formalization time. The state's name is to readily understand the workflow and the blocks connected with the \textit{ML\_Block\_Connection} stereotype.

As the scope of this work is to formalize the machine learning and not to improve the executable code or to derive the code automatically, the result of the machine learning and the implementation itself are not depicted and left to future work.

\subsection{UC2 - 3D Printer Success Evaluation during Printing}
The purpose of the application is to detect faulty 3D prints during the printing process by comparing the actual status of the printed model with the intended model.
This use case illustrates the method's applicability to other data sources, such as image data, and the integration of the method into an executable workflow engine.
Additionally, the integration of pre-trained models is depicted by integrating TensorFlow Hub.
The idea of image similarity is based on an image similarity tutorial\footnote{\url{https://towardsdatascience.com/image-similarity-with-deep-learning-c17d83068f59}}.

\begin{figure}
    \centering
    \includegraphics[width=\textwidth]{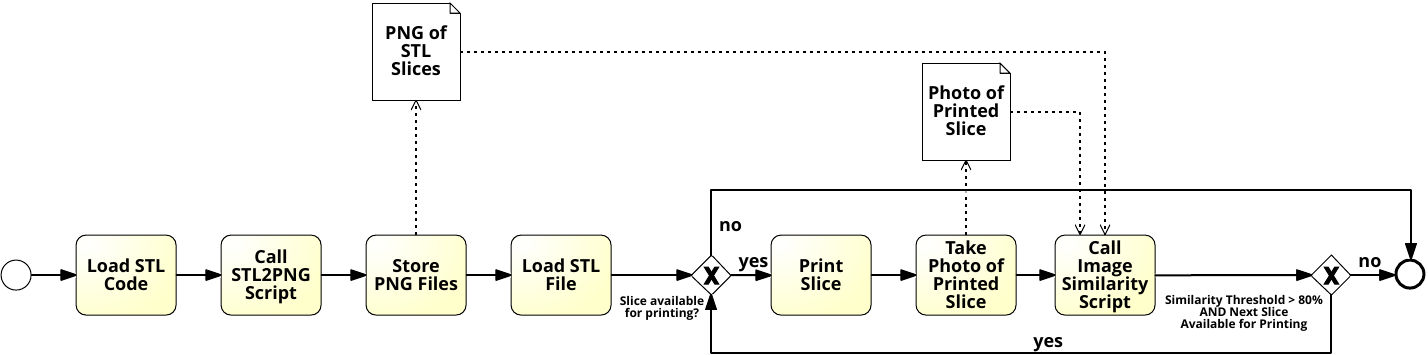}
    \caption{Workflow Integrating the formalized machine learning method to early abort 3D printing.}
    \label{fig:UC2_3DPrinter}
\end{figure}

The use case process is described below and illustrated in Figure \ref{fig:UC2_3DPrinter}.
We adopt the CPEE process engine \citep{mangler_cpee_2014,mangler_cloud_2022} to orchestrate the application process, as the CPEE provides a lightweight and straightforward user interface to orchestrate any application that allows interaction via REST web services.
Figure \ref{fig:UC2_3DPrinter} shows the workflow of the application, consisting of image generation and printing.
The first three process steps define the slicing of a STL file and the generation of the reference images.
Particularly, a Python script is called that generates the slices based on a given STL file and stores the generated reference images for later comparison and similarity check.
The second part of the process consists of a loop that prints a slice, takes a photo with a camera from the top center of the working area, and calls a similarity script to compare the intended and actual printed model.
The image similarity algorithm is defined using the machine learning formalization method, proposed here.
The defined algorithm provides a similarity index compared to a threshold value.
If the threshold is exceeded, the printing process is aborted, otherwise, it is repeated.

\begin{figure}
    \centering
    \includegraphics[width=0.5\textwidth]{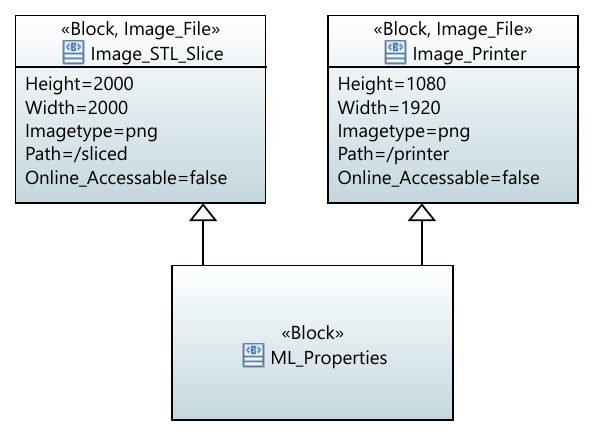}
    \caption{Image definition used for the similarity prediction.}
    \label{fig:UC2_01_Input}
\end{figure}

The machine learning model integrated into the printing process is formalized below.
Figure \ref{fig:UC2_01_Input} shows input data consisting of two images: the image sliced from the STL file and the photo from the 3D printer camera.
In contrast to the first use case, the data attributes are not further detailed with stereotypes because the input data do not show any variations, i.e. the format and resolution of the images do not change.

\begin{figure}
    \centering
    \includegraphics[width=\textwidth]{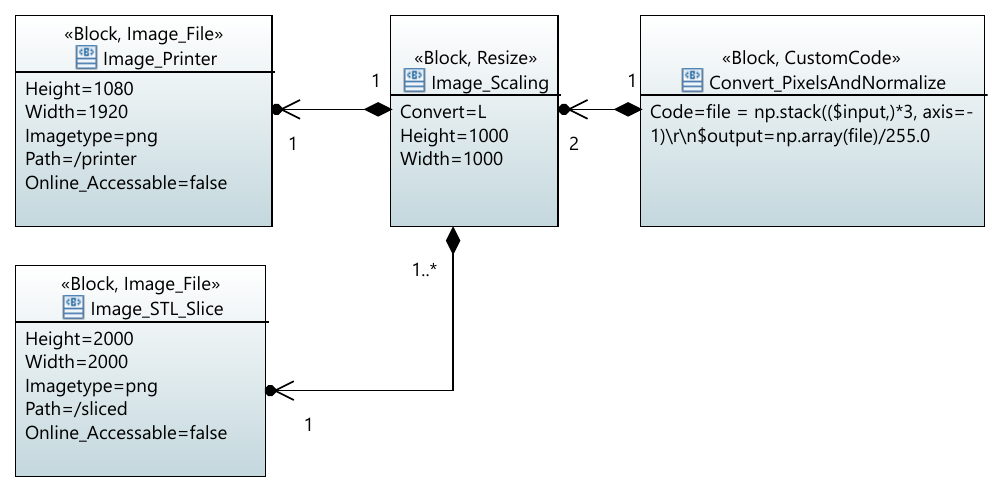}
    \caption{Image scaling and normalization used for data preprocessing.}
    \label{fig:UC2_02_Preprocessing}
\end{figure}

Figure \ref{fig:UC2_02_Preprocessing} depicts the scaling of the images such that they have the same dimension.
The conversion parameter \textit{L} allows comparing the images on a black-and-white basis.
Normalization of the pixels and colors between 0 and 1 is also applied.
The normalization in the block \textit{Convert\_PixelsAndNormalize} should be defined as a new stereotype.
In this case, we show the application of the \textit{CustomCode} stereotype, allowing for the injection of program code, which allows rapid prototyping.
However, flaws, such as vulnerability or hijacking of the method might lead to reduced understanding and reproducibility.
Additionally, it is not the purpose of the method to insert programmed code.
For further discussion, see Section \ref{sec:discussion_potential}.

\begin{figure}
    \centering
    \includegraphics[width=\textwidth]{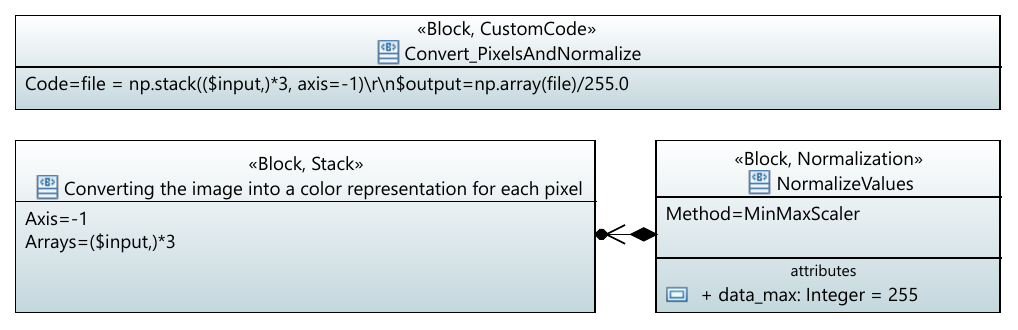}
    \caption{On top the wrong application of the method and below correct use.}
    \label{fig:UC2_02_Preprocessing_correct}
\end{figure}

With respect to potentially wrong use of the method, Figure \ref{fig:UC2_02_Preprocessing_correct} depicts the wrong used stereotype \textit{CustomCode} on top and below the correct use of stereotypes for the same result with a slightly changed code sequence.

\begin{figure}
    \centering
    \includegraphics[width=\textwidth]{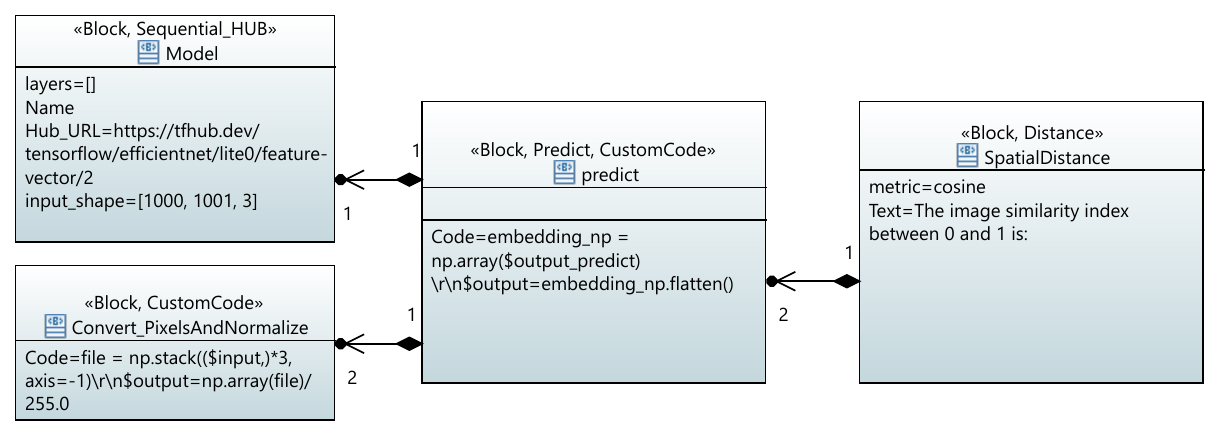}
    \caption{Integration of pre-trained model and prediction with cosine distance to express the similarity of the images.}
    \label{fig:UC2_03_Algorithm}
\end{figure}

Further, the two images are fed to the classification algorithm, as illustrated in Figure \ref{fig:UC2_03_Algorithm}.
The input value \textit{Model} describes a TensorFlow Hub input, a pre-trained model to classify images.
Finally, the result is measured using \textit{cosine} distance metrics.
The threshold for canceling the printing is implemented in the workflow and can be adjusted by the user.

\begin{figure}
    \centering
    \includegraphics[width=\textwidth]{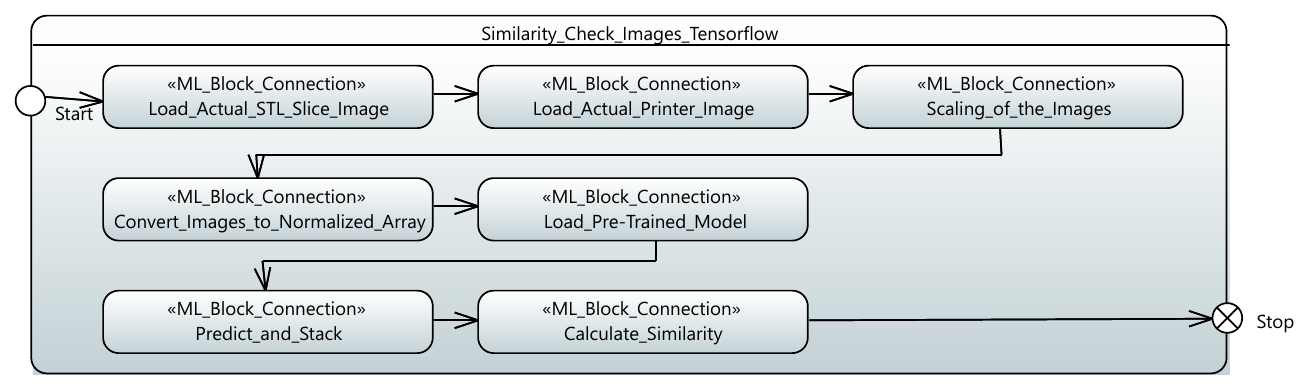}
    \caption{The execution workflow of the TensorFlow-based prediction algorithm.}
    \label{fig:UC2_04_Workflow}
\end{figure}

Finally, Figure \ref{fig:UC2_04_Workflow} depicts the execution sequence of the algorithm.

\section{User Study}\label{sec:userstudy}
Typical user of the presented method are computer scientists and engineers from various disciplines, depending on the application area.
Therefore, this study aims to assess and compare computer scientists' and mechanical engineers' subjective workload and user experience regarding understanding, modifying, and creating machine learning functions in a model-based method.
Further, the time required for applying changes or creating constructs in SysML is assessed to allow a comparison of the participants based on previous experiences, e.g., programming or modeling prior knowledge.
Since the study and the modeling is conducted using the SysML modeling tool Papyrus\footnote{\url{https://www.eclipse.org/papyrus/}}, it is impossible to eliminate distortions due to the usability of the underlying tool, e.g., 
``How to model a block''.
Therefore, the study director will provide verbal assistance if a participant requires support due to the tool's usability.

Large sample sizes are necessary to enable quantitative evaluation, which is not applicable due to resource constraints.
Therefore, the principles of discount usability are applied to test only a small group of customers and to identify the main usability problems by applying small qualitative user studies with three to five users, a detailed scenario, and a think-aloud method \cite{nielsen_usability_1993}.
According to \cite{nielsen_usability_1993}, a 70\% chance to find 80\% of the usability issues is given with five users.
However, in literature, there are reports that the increase of five participants to ten significantly changes the amount of found issues \cite{faulkner_beyond_2003}.
In this respect, a total number of 12 users were tested, equally distributed among the two groups, Computer Scientists (CS) and Mechanical Engineers (ME).

In the following, the experimental setting is illustrated.
Next, an introduction to the evaluation procedure is given, followed by an introduction of the test cases in Section \ref{sec:userstudytestcases}.
Finally, the results of the user studies are depicted in Section \ref{sec:surveyResults}.
A discussion on the implications from the user study is given in Section \ref{sec:userstudydiscussion}.

\subsection{Experimental Setting}
The user study was conducted with 12 participants.
Each participant has a university degree (B.Sc., M.Sc., or Ph.D.) and received a basic introduction to programming at university.
Half of the participants are CSs, and half MEs.
Other engineers can serve as potential users and equally valid test users, as well.
However, to obtain a more homogeneous group, engineers are limited to MEs.

Due to the participants' different knowledge in modeling, programming, and data science, a self-assessment of their experience was made at the beginning of the user test.
Table \ref{tab:users} summarizes the knowledge levels of the participants based on their highest university degree, years of experience, position at the current job, and self-assessment on the three relevant dimensions.

\begin{table}
\caption{Participants of the user study aligned with self-assessment of experience.}
\label{tab:users}
\begin{tabular}{@{}lllllll@{}}
\toprule
User & \begin{tabular}[c]{@{}l@{}}Univ.\\ Degree\end{tabular} & \begin{tabular}[c]{@{}l@{}}Years of\\Experience\end{tabular} & Position & \begin{tabular}[c]{@{}l@{}}Programming \\ Skills\end{tabular} & \begin{tabular}[c]{@{}l@{}}Data Science\\Skills\end{tabular} & \begin{tabular}[c]{@{}l@{}}UML/\\SysML\\Skills\end{tabular} \\ \midrule
 CS-1 & B.Sc. & 5 & \begin{tabular}[c]{@{}l@{}}Software\\Engineer\end{tabular} & 7 & 3 & 6 \\ 
 CS-2 & M.Sc. & 3 & \begin{tabular}[c]{@{}l@{}}Software\\Engineer\end{tabular} & 8 & 6 & 7 \\ 
 CS-3 & M.Sc. & 1 & \begin{tabular}[c]{@{}l@{}}Ph.D.\\Student\end{tabular} & 7 & 6 & 3 \\ 
 CS-4 & M.Sc. & 2 & \begin{tabular}[c]{@{}l@{}}Ph.D.\\Student\end{tabular} & 6 & 7 & 6 \\ 
 CS-5 & M.Sc. & 1 & \begin{tabular}[c]{@{}l@{}}Ph.D.\\Student\end{tabular} & 6 & 7 & 8 \\ 
 CS-6 & B.Sc. & 1 & \begin{tabular}[c]{@{}l@{}}Application\\Manager\end{tabular} & 7 & 4 & 4 \\ 
 ME-1 & M.Sc. & 6 & \begin{tabular}[c]{@{}l@{}}Project\\Manager\end{tabular} & 4 & 1 & 2 \\ 
 ME-2 & B.Sc. & 11 & \begin{tabular}[c]{@{}l@{}}Project\\Manager\end{tabular} & 2 & 3 & 1 \\ 
 ME-3 & Ph.D. & 10 & \begin{tabular}[c]{@{}l@{}}Digital\\Engineering\\Manager\end{tabular} & 6 & 4 & 8 \\ 
 ME-4 & B.Sc. & 2 & \begin{tabular}[c]{@{}l@{}}Simulation\\Engineer\end{tabular} & 2 & 2 & 1 \\ 
 ME-5 & M.Sc. & 3 & \begin{tabular}[c]{@{}l@{}}Expert\\Powertrain\end{tabular} & 2 & 1 & 3 \\ 
 ME-6 & M.Sc. & 1 & \begin{tabular}[c]{@{}l@{}}Manufacturing\\Engineer\end{tabular} & 1 & 2 & 1 \\ \bottomrule 
\end{tabular}
\end{table}

\subsection{Evaluation Procedure}
The study started with a basic introduction to SysML and an overview of the method introduced in this work, taking approximately 10 minutes and involving the presentation of two predefined block definition diagrams as samples with a focus on the modeling and understanding of a block definition diagram and the application of the introduced stereotypes.

Following this, the users had to perform three tasks, i.e., 
(1) showing that they understand the purpose of the modeling and the basic idea of the method by describing the modeled methods in Figure \ref{fig:implPreProcessing}, (2) replacing a \textit{CSV} stereotype with \textit{Text-file} stereotype and redefining the attribute properties of the text file, and (3) adding a new function by connecting a new block with a particular stereotype to an existing block.

Each of the tasks (1) -- (3) is subdivided into sub-activities to allow fine-grained evaluation of the tasks and the performance achieved by the participants.
The sub-activities are presented with their tasks in Table \ref{tab:USSubtasks}.

\begin{table}
    \centering
    \caption{The three main tasks to be performed by the participants, with subtasks that can be used to assess whether the task has been completed.}
    \label{tab:USSubtasks}
    \begin{tabular}{l|l}
\textbf{Main Task} & \textbf{Subtask}\\\hline
\textbf{Task 1 Understanding} & Identification of input files\\
 & Description of values stored in CSV\_2 input file\\
 & Description of attributes of the data stereotype of CSV\_2 values\\
 & Identification of stereotype properties, e.g. path of CSV\_2 file\\
 & \\
\textbf{Task 2 Changes} & Stereotype identified\\
 & Stereotype removed\\
 & Stereotype added\\
 & Stereotype attribute identified\\
 & Stereotype attribute value set\\
 & \\
\textbf{Task 3 Modeling} & Block added to view\\
 & Block associated with input\\
 & Stereotype added\\
 & Stereotype attribute value set\\
    \end{tabular}
\end{table}

For each participant, the time taken to perform the tasks is recorded.
After each of the three tasks, NASA Task Load Index (NASA-TLX, \cite{hart_development_1988,hart_nasa-task_2006}) and the Systems Usability Scale (SUS, \cite{brooke_sus_1996}) questionnaire are filled out by the users to assess the participants' subjective workload and usability.
Before filling out the questionnaire, the users were explicitly told to evaluate the method's usability, not Papyrus's.

\subsection{Test Cases}\label{sec:userstudytestcases}
Table \ref{tab:USSubtasks} depicts the subtasks to accomplish the tasks of the user study. 
Therefore, each subtask is assessed by the study leader to determine whether they are completed correctly or not.
If a user could not find a specific button due to the usability of Papyrus, but could justify why it is being searched for, e.g., ``I need to remove a stereotype and add a new one so that a new function is defined'', the task is evaluated as correct.

To achieve reproducibility, the tasks were set exactly with the following wordings:

\begin{enumerate}[align=left]
    \item[Task 1 Understanding:] Please describe what can be seen in the currently displayed diagram and what function it fulfills. Additionally, please answer the following questions:
    \begin{enumerate}
        \item What are the two input files, and in which format?
        \item What values are stored within CSV\_2?
        \item What is the type of date\_date, and how is it represented in the ML model?
        \item What are the path and encoding of the two input files?
        \item What are the properties of DataFrame\_Merge Stereotype?
    \end{enumerate}
    
    \item[Task 2 Function Exchange:] Behind the here presented \textit{TextFile} function, a \textit{CSV} stereotype is defined.
    However, the type is incorrect. 
    Please change the file type to \textbf{\textit{Text-File}}.
    Additionally, set the encoding to \textbf{\textit{UTF-8}} and the path to \textbf{\textit{C:/file.txt}}.
    
    \item[Task 3 Adding a Function:]In the following view, you can see two input files connected to a merge block.
    Additionally, a normalization of the merge block is required.
    Please add the function for \textbf{\textit{Normalization}} and set the value of the normalization method to \textbf{\textit{MaxAbsScalar}}.
\end{enumerate}

\subsection{Survey Results}\label{sec:surveyResults}
\begin{figure}
    \centering
    \includegraphics[width=\textwidth]{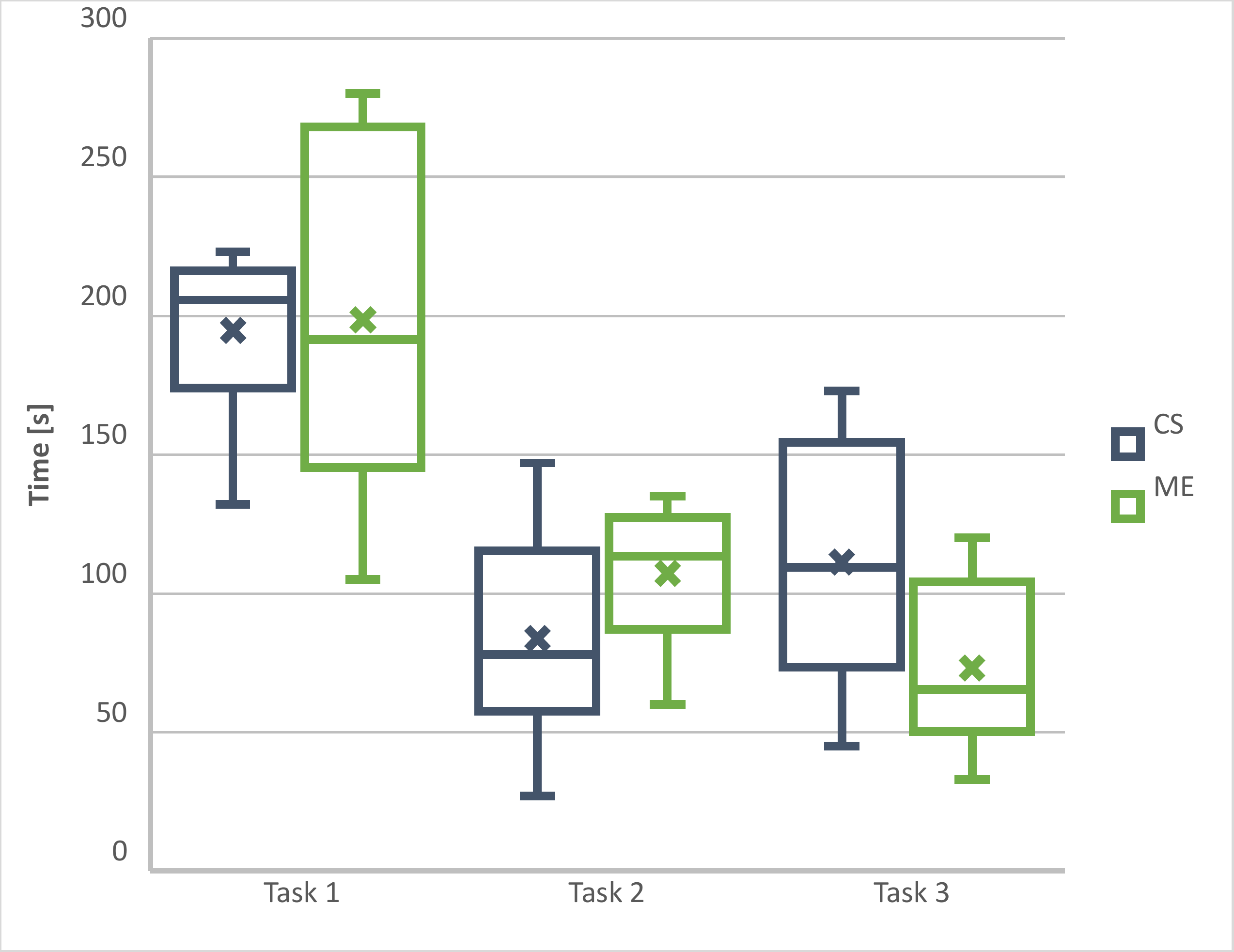}
    \caption{The time required by the participants per task and training direction.}
    \label{fig:Time}
\end{figure}

Figure \ref{fig:Time} shows boxplots of the required times for the individual tasks grouped per task and training of the participants in CS or ME.

For Task1, the time required is higher than for Task2 and Task3, whereas Task2 and Task3 shows a comparable average and distribution.
One reason for the higher time for Task1 is that the users had to describe a model and this task is therefore more time-consuming.
It was also observed that repetitive tasks made the users faster, which also came as feedback from the participants.
Further, the dispersion of Task1 for ME is higher compared to CS.
This scatter might be explained because of the varying experience levels of the participants with respect to modeling and data science.
However, there was no correlation between the time spent and the correctness of the execution of the sub-activities.
Regarding the dispersion of CS, interestingly, Tasks 2 and 3 vary more than Task1.
This can mainly be explained by the familiarity with the Papyrus modeling environment.
Thus, participants with more Papyrus experience had completed the tasks much faster than those who used Papyrus for the first time.

\begin{figure}
    \centering
    \includegraphics[width=\textwidth]{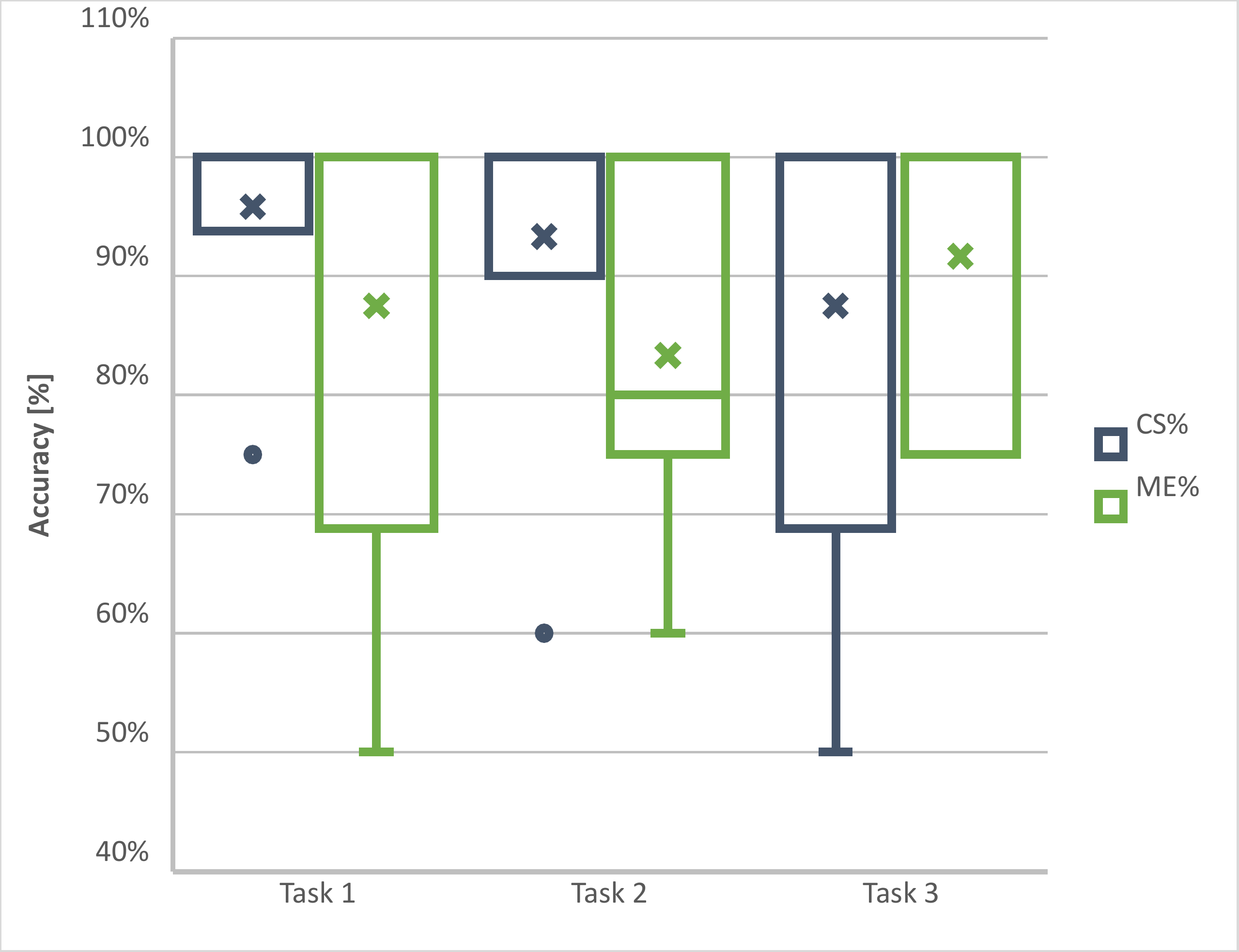}
    \caption{The degree of correct performance of the tasks.}
    \label{fig:Accuracy}
\end{figure}

Figure \ref{fig:Accuracy} shows the result of the individual tasks in terms of correctness in relation to the subtasks of Table \ref{tab:USSubtasks}.
CS perform better for T1 and T2, which can be explained by the extended prior experience regarding UML of CS obtained during university education.
In T3, however, ME perform better.
This can be explained by an outlier value for CS that performs significantly below the average.
The overall accuracy of ME increased with the evolving tasks although the average of T2 is lower than for T1.

\begin{figure}
    \centering
    \includegraphics[width=\textwidth]{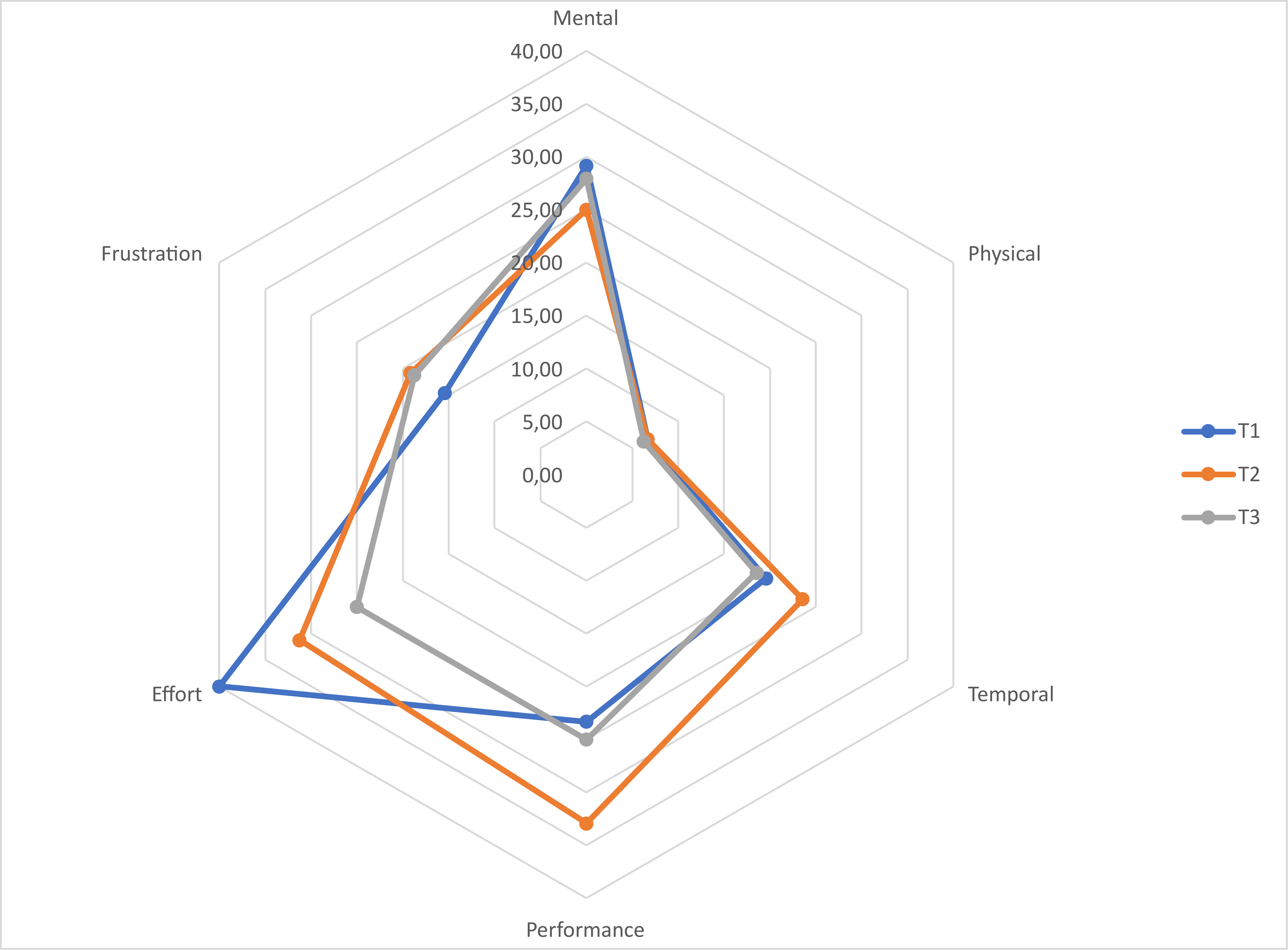}
    \caption{Result of the NASA-TLX questionnaire.}
    \label{fig:nasa-tlx}
\end{figure}

The results of the applied NASA-TLX test to indicate the perceived workload of the participants for the specific tasks are presented in Figure \ref{fig:nasa-tlx}.
The lower the value of a dimension of the NASA-TLX, the lower the perceived workload.
Consequently, a low scale value is seen as positive.
The \textit{Effort} dimension shows, for example, that with increasing experience or task, the perceived effort decreases.
Further, the frustration increases and the performance decreases compared to T1.
For T3, the standard error is larger than for T1 and T2.
Both might be justified due to the increasing complexity of the tasks.
However, it is a contrast compared to the achieved accuracy in Figure \ref{fig:Accuracy}.

\begin{figure}
    \centering
    \includegraphics[width=\textwidth]{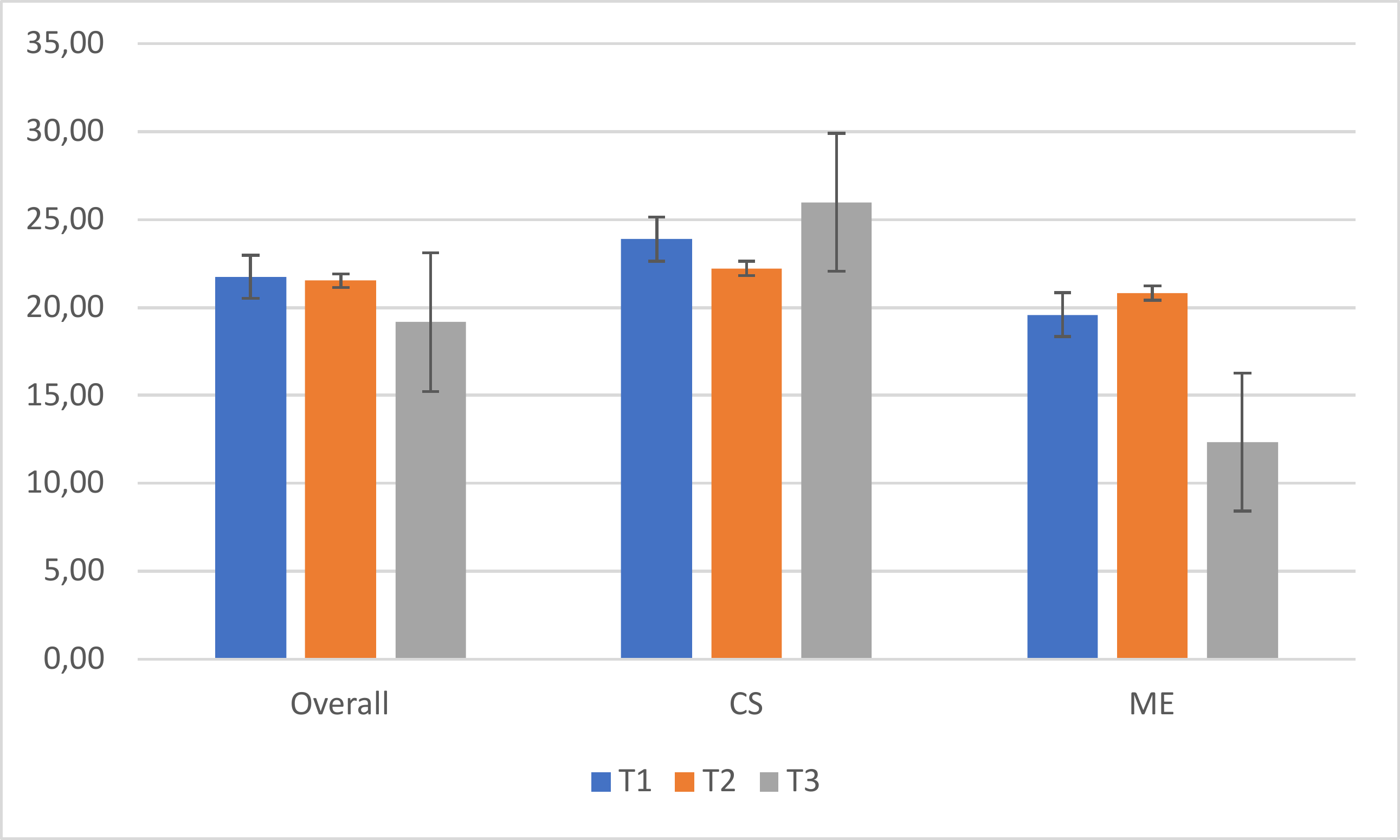}
    \caption{NASA-TLX overall score.}
    \label{fig:nasa-tlx-overall}
\end{figure}

The raw overall scores of the tasks are depicted in Table \ref{fig:nasa-tlx-overall}.
According to \cite{hancock_human_1988,prabaswari_mental_2019}, the workload is categorized as `medium', which is the second best score and ranging from 10 to 29 points.
The cumulative results of CS and ME shows a decreasing workload among the evolving tasks.
For CS, the workload appears to be higher than for ME, especially for T3.
As of the user feedback, no justification can be given on the difference between CS and ME.

The results of the SUS test with different rating scales are shown in Table \ref{tab:sus} based on \cite{blattgerste_web-based_2022}.

\begin{landscape}
\begin{table}
    \centering
    \caption{SUS analysis results.}\label{tab:sus}
    \begin{tabular}{|l|l|l|l|l|l|l|l|l|l|l|l|} 
    \hline
    \textbf{Variable} & \begin{tabular}[c]{@{}l@{}}\textbf{SUS }\\\textbf{Score }\\\textbf{(mean)}\end{tabular} & \textbf{Percentile} & \textbf{SD} & \textbf{Min} & \textbf{Max} & \textbf{1. Quartile} & \textbf{Median} & \textbf{3. Quartile} & \begin{tabular}[c]{@{}l@{}}\textbf{Adjective }\\\textbf{Scale}\end{tabular} & \begin{tabular}[c]{@{}l@{}}\textbf{Quartile }\\\textbf{Scale}\end{tabular} & \begin{tabular}[c]{@{}l@{}}\textbf{Acceptability }\\\textbf{Scale}\end{tabular} \\ 
    \hline
    \textbf{T1 - CS} & 75.0 & 72.77 & 10.7 & 60.0 & 92.5 & 67.5 & 71.25 & 86.875 & Good & 3rd & Acceptable \\ 
    \hline
    \textbf{T1 - ME} & 71.25 & 60.08 & 7.03 & 62.5 & 82.5 & 64.375 & 70.0 & 78.75 & Good & 3rd & Marginal \\ 
    \hline
    \textbf{T2 - CS} & 72.5 & 64.38 & 18.65 & 37.5 & 92.5 & 56.25 & 76.25 & 90.625 & Good & 3rd & Marginal \\ 
    \hline
    \textbf{T2 - ME} & 71.25 & 60.08 & 16.12 & 50.0 & 97.5 & 55.625 & 68.75 & 88.125 & Good & 3rd & Marginal \\ 
    \hline
    \textbf{T3 - CS} & 72.08 & 62.95 & 13.8 & 47.5 & 92.5 & 60.625 & 73.75 & 83.125 & Good & 3rd & Marginal \\ 
    \hline
    \textbf{T3 - ME} & 77.08 & 79.24 & 13.1 & 62.5 & 97.5 & 62.5 & 75.0 & 91.875 & Good & 3rd & Acceptable \\
    \hline
    \end{tabular}
\end{table}
\end{landscape}

\begin{figure}
    \centering
    \includegraphics[width=\textwidth]{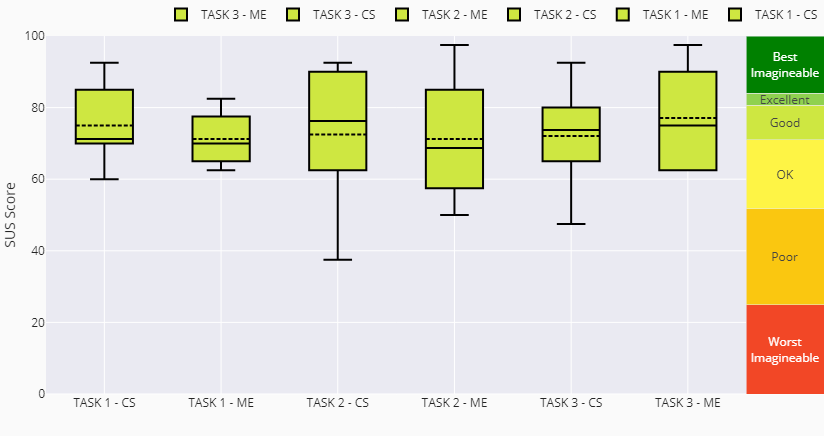}
    \caption{Boxplot of the SUS score.}
    \label{fig:susboxplot}
\end{figure}

Figure \ref{fig:susboxplot} presents the SUS score as a boxplot, prepared with an online tool for analyzing SUS questionnaire results \cite{blattgerste_web-based_2022}.

The \textit{adjective scale} score in the boxplot is aligned with \cite{jeff_5_2018}, which is based on \cite{bangor_empirical_2008}.
The figure highlights that each task achieves the rating good for both CS and ME.
The standard error of CS is slightly higher than for ME, which can also be seen in Table \ref{tab:sus}.
The values of quartile scale shown in Table \ref{tab:sus} are according to \cite{bangor_determining_2009} and acceptability scale according to \cite{bangor_empirical_2008}.
ME increased the score in T3, T1 and T2 are equal.
CS decreased the score among the tasks.
However, the changes in the scores are little and therefore not justifiable.

\begin{figure}
    \centering
    \includegraphics[width=\textwidth]{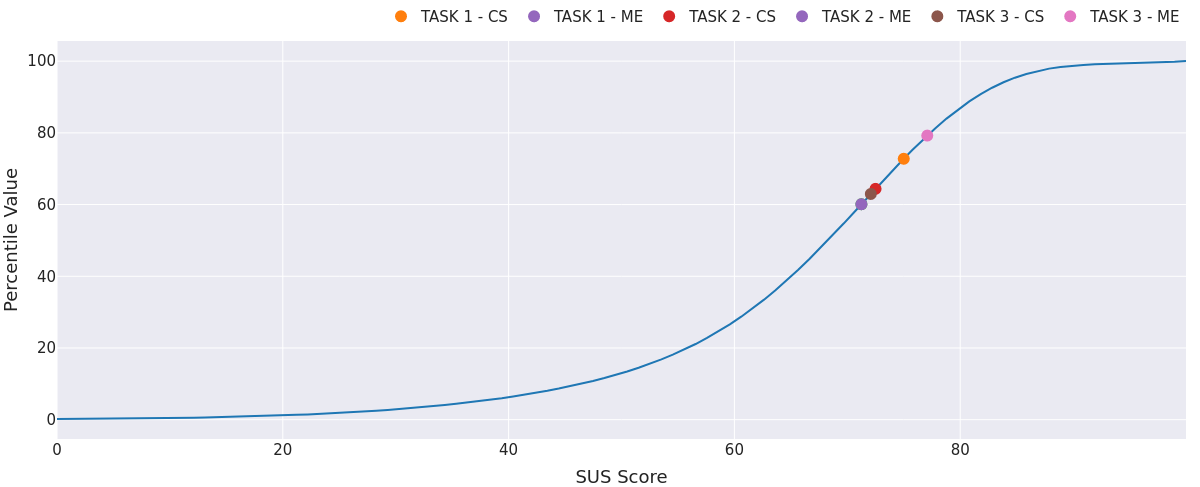}
    \caption{Percentile curve of the SUS questionnaire.}
    \label{fig:suspercentile}
\end{figure}

Figure \ref{fig:suspercentile} depicts the percentile scale based on \cite{sauro_quantifying_2016}.
Since the percentile score is not uniform or normally distributed, a percentile score was created based on 5000 SUS studies.
In this respect, the comparison shows that the tests achieved a percentile between 60 and 79.
T3 ME over performed with 79.
For CS and ME the average percentile is ~66.
T1 and T2 for ME have exactly the same value, which is why they are shown as one colour in the Figure.

\section{Discussion}\label{sec:discussion}
This section discusses advantages and potential flaws of the newly introduced method to formalize machine learning tasks. The structure of the section is as follows: First, the metamodel's extension and the stereotypes' proposed structure are discussed. Next, the benefits and shortcomings of the modeling semantic are assessed with a particular focus on the applicability and potential ambiguous interpretation. Next, potential risks of model-driven machine learning and future work are presented. Finally, the implications of the user study are presented and discussed.

\subsection{Stereotypes and Structure of the Custom Metamodel}
The integration of custom stereotypes has been proven beneficial in the literature \citep{kuzniarz_empirical_2004}. In this method, the use of stereotypes to encapsulate and abstract knowledge about machine learning tasks is beneficial as implementation details are hidden, thus supporting communication between different engineers not necessarily experienced in machine learning or programming.
With structuring the stereotypes using packages, a stereotype organization aligned to the CRISP-DM methodology is given, supporting refinements and extension in a fine-grained, hierarchical manner. Particularly, the definition of blackbox and abstract stereotypes allows the description of various functions without the necessity to specify each machine learning function in detail. 
In the custom metamodel, custom \textit{Enumerations} are defined to limit the number of attribute values, which reduces the model's wrong specifications. Another opportunity to reduce the scope of possible selections is to reduce the number of allowed stereotypes, e.g., only inheritance of the abstract stereotype \textit{PreProcessing} can be assigned as a value for a specific attribute.
However, the filtering of stereotypes requires specific rules that have not yet been integrated or elaborated.
Although various methods are defined using stereotypes, the level of detail might be too little for practical application. \textit{DateConversion}, for example, can be applied to manifold input values and various outputs, e.g., output representation as a string or Coordinated Universal Time (UTC). Adding multiple \textit{DateConversion} stereotypes for each case is possible. Still, with a growing number of stereotypes, the complexity of selecting the correct, unambiguous stereotype increases while the maintainability decreases. Similarly, if too many stereotype attributes have to be set, the complexity and the effort for the application increases.
With respect to these uncertainties at the level of detail required for fine-grained definition of machine learning tasks, industrial case studies have to be conducted to elaborate and validate sufficient degree of detail and additionally define future work.

\subsection{Complexity of Unambiguous Modeling}
The definition of an implementation structure aligned with the CRISP-DM methodology starting from the business understanding and ending with the definition of evaluation and workflows, is promising to be useful due to the integration of a comprehensive and mature methodology in a MBSE method. Additionally, more experienced computer scientists aware of CRISP-DM can rely on experiences and the benefits of CRISP-DM. Furthermore, in practice, one third of data scientists lack business understanding and communication skills\citep{anaconda_state_2022}, which can be supported by the model-based method of CRISP-DM.\newline
Each block implementing a \textit{ML} stereotype within the implementation structure can be seen as an encapsulated subtask. Each subtask provides an output that can be used as input for another block. However, the given method does not explicitly specify the output of a block. Therefore, the output is defined by the implementing computer scientist, which may lead to different results due to the range of experience of the decisions and the laziness of the semantics, which allows to create arbitrary associations that may not be implementable.
In this respect, future work requires the integration of model checking to reduce orphan associations, infeasible implementations and unwanted side effects on changing associations. 

Despite the ambitiousness of the modeling and the potential errors of the associations, the method supports the elaboration and definition of machine learning tasks from early development, which is beneficial. The authors believe that the flaws in the beginning of the method are getting less with the application due to the possibility of reusing certain parts of the formalization. The reuse additionally allows to preserve knowledge and contribute to standardization in the modeling and implementation, which further leads to a reduction of cost and risk in the design \citep{beihoff_world_2014} and the maintenance of machine learning applications.

\subsection{Potential of Model-Driven Machine Learning}\label{sec:discussion_potential}
The given proposal to describe machine learning tasks using a model-based method has some benefits but also disadvantages.
A core disadvantage is the initial effort to introduce stereotypes and formalize the model. 
In this respect, traditional programming might be less time consuming and therefore, users might use the \textit{CustomCode} stereotype to inject code.
However, it is not the purpose of the method to insert code injection due to vulnerability risks and the reduced documentation and understanding by others.
Consequently, future work is required to investigate an extension of the method that allows to generate code from the model but with limitations so that code injections like described in the use case are not possible.
Another disadvantage of the stereotypes is the potential effort for maintenance if interfaces are proprietary or rapidly changing, e.g. due to configuration changes or replacement of machines.
Closely related, for huge projects, the complexity of the resulting models might be very high, including potential errors in the model or ambiguous associations, which might be very hard to find and thus lead to additional communication effort.
Nevertheless, the shortcoming of a complex ramp-up might also be a benefit in the end due to the possibility of introducing model libraries containing well-defined models, leading to standardized parts that can be reused. Further, the method allows to use the formalization as documentation of the implemented technologies that improve the maintainability and extendability for various engineers. Additionally, with further investigations regarding model validation and model debugging features, errors in the semantics can be found and repaired without actually implementing the machine learning application. However, to use this efficiently, the integration into advanced model lifecycle management \citep{fisher_311_2014} might be necessary to allow collaborative working.\newline
Due to the non-programming description of machine learning, the method is promising to increase the communication among various disciplines. In particular, with the integration of the general-purpose language SysML and the intersection of CRISP-DM and MBSE, the heterogeneous communities are broadly supported, which favors the implementation of machine learning in industrial practice and supports to shift knowledge in enterprises regarding machine learning. Further, the method can be integrated into early product development due to the abstract definition that allows to foresee various data interfaces which might have been forgotten during the development. This potentially leads to increased accuracy of the machine learning applications and might reduce failing machine learning projects, which is a well-known problem in industries \citep{radler_survey_2022}.
In this section, the advantages and potential shortcomings of the method have been shown. However, the key advantages of formalized knowledge was not detailed yet. The machine-readable artifacts (models) are usable with model transformations so to generate executable code, such as a Python script. 
Particularly, each \textit{ML} stereotype consists of knowledge to describe a specific subtask, which is a function in a programming language, e.g. a date conversion. The function parameters are defined in the stereotype (mandatory parameters) or on the block (optional parameters). Since stereotypes have to be uniquely named, each can be mapped to a generic code template in a dedicated programming language, e.g. Python. The templates consist of fixed code and generic parts with placeholders, which are filled based on the model's attributes. The state diagram defines the execution order; all blocks are a well-encapsulated functionality; hence, each block can generate a single code block in an Jupyter Notebook\footnote{\url{https://ipython.org/notebook.html}}. With the automatic derivation of executable machine learning code, the effort for the documentation and implementation is reduced and potentially lead to less errors in the interpretation.
In this respect, future work consists of implementing a proof of concept showing that a derivation and decomposition of formalized machine learning knowledge is beneficial.

\subsection{Implications from the User Study}\label{sec:userstudydiscussion}
The user study was conducted with two groups that are representative for using the method presented in this work in practice.
The results show that the majority of the tasks were successfully accomplished.
From a study perspective, the users could perform each task without additional guidance on the modeling method.
Still, problems occurred with the user-interface of Papyrus, e.g., expanding a group of elements to select a \textit{block} element for modeling. However, learning effects could be observed among the tasks on both CS and ME.

The assessment of the NASA-TLX showed that the mental demand for each task is comparable.
A similar observation can be made for the level of frustration, which is slightly lower for the first task.
Contrary to expectations, the participants perceived the effort as decreasing. 
With regard to the task, the effort for modeling should have been higher than for understanding a model.
Nevertheless, it can be implied that both CS and ME can use the method in terms of task load without being more strained.

From an usability perspective, the method achieved good results.
Users rated especially the consistency of the method as very high.
Comparing the method with others using the percentile curve, it achieved a rank over 66.

However, the first positive results could be due to some shortcomings in the study design.
In particular, the demand for rating Papyrus might have a larger impact on the study design than expected.
The usability feeling of the users is more dedicated to the experience with Papyrus than to the method, although it was said before to focus on the method.
In this respect, a paper prototype where users had to move paper snippets on the table might have been more valuable.
Furthermore, most of the participants reported their data science knowledge as low and yet were able to explain what happens in a given model or create a model building block themselves.
However, modeling their own data science application might not be possible, as the general understanding of data science is too low.

Nevertheless, it can be seen as a result of the study that the modeled knowledge can be used as a communication medium.
Therefore, it should also be possible for non-data scientists to perform a plausibility analysis, as they can gain an understanding of the process without understanding programming code.

However, this would need to be evaluated in a further study.
Similarly, an evaluation of the results with the help of a larger study should be sought. 
\section{Conclusions}\label{sec:conclusion}
In this work machine learning task definition using means of SysML is depicted. Particularly, the metamodel of SysML is extended with stereotypes to reflect functions from the machine learning domain. Additionally, the CRISP-DM methodology is used as basis for the structure of the models to organize the development with specific viewpoints. 
The method is evaluated in a case study showing the integration of machine learning task definition in a cyber-physical system as well as in a case study where a workflow engine is integrated for the interruption of a 3D printer task if the aimed result cannot be achieved.
Additionally, a user study is performed to collect an overview of the perceived workload using NASA-TLX questionnaire and to check usability of the system using the SUS questionnaire.
The findings of the evaluation showed that the entire workflow of a machine learning solution can be reflected using SysML. Additionally, the connection between the domain of (mechanical/electrical) engineers and machine learning experts is shown.
With the MBSE integration and the involvement of various stakeholders from different disciplines, an improvement in communication is expected as shown in a user study. 
The user study implies that non-experts in data science can use the method as medium of communication.
Future work consists of the extension of the method to automatically derive executable machine learning code acting as a basis for the implementation. In addition, a case study must be conducted to develop a minimum level of detail required to sufficiently define a machine learning model that can be used for communication, and thus guide the implementation of the executable code through the formalization of the machine learning model.

\bibliography{references}


\end{document}